\documentclass[twoside]{article}
\usepackage{graphicx}
\usepackage{array}
\usepackage{url}
\usepackage{lipsum} 
\usepackage[sc]{mathpazo} 
\usepackage[T1]{fontenc} 
\usepackage{microtype} 
\usepackage[hmarginratio=1:1,top=32mm,columnsep=20pt]{geometry} 
\usepackage{multicol} 
\usepackage[hang, small,labelfont=bf,up,textfont=it,up]{caption} 
\usepackage{booktabs} 
\usepackage{float} 
\usepackage{hyperref} 
\usepackage{lettrine} 
\usepackage{paralist} 
\usepackage{abstract} 
\usepackage{color}

\usepackage{titlesec} 
\usepackage{fancyhdr} 
\title{\vspace{-15mm}\fontsize{24pt}{10pt}\selectfont\textbf{Simulation of High Density Pedestrian Flow: \\ Microscopic Model}} 

\author{
\large
\textsc{Mohamed H. Dridi}\\
\normalsize Unversit{\"a}t  Stuttgart \\
1. Institute f{\"u}r Theoretische Physik \\
\normalsize \href{mailto:mohamed.dridi@itp1.uni-stuttgart.de}{mohamed.dridi@itp1.uni-stuttgart.de} 
\vspace{5mm}}

\date{26/01/2015} 

\begin{document}
\maketitle 
\thispagestyle{fancy}

\begin{abstract}
\noindent  In recent years modelling crowd and evacuation dynamics has become very important, with increasing huge numbers of people gathering around the world for many reasons and events. The fact that our global population grows dramatically every year and the current public transport systems are able to transport large amounts of people, heightens the risk of crowd panic or crush. Pedestrian models are based on macroscopic or microscopic behaviour. In this paper, we are interested in developing models that can be used for evacuation control strategies.
This model will be based on microscopic pedestrian simulation models, and its evolution and design requires a lot of information and data. The people stream will be simulated, based on mathematical models derived from empirical data about pedestrian flows. 
This model is developed from image data bases, so called empirical data, taken from a video camera or data obtained using human detectors. We consider the individuals as autonomous particles interacting through social and physical forces, which is one approach that has been used to simulate crowd behaviour.

Typical questions in pedestrian motion planning and design are: How many minutes would be needed to evacuate a football stadium or a pop concert arena? Where is the best place for pedestrian information or a poster wall to be placed? How many exit signs should be distributed over the area and where should they be placed? How many minutes does a pedestrian need to go from point $A$ to point $B$ in the overcrowded places? Are transfer times still realistic when the pedestrians are not familiar with the layout?. For better planning in this area a greater understanding of human behaviour is required and one way to achieve this is to improve the tools available for pedestrian planning (such as pedestrian micro simulation).
In particular, particle-based simulation has become an attractive paradigm for modelling and simulating pedestrian decision-making and behaviour because it supports a one-to-one correspondence between the subject of observation and the simulated particle.

The target of this work is to describe a comprehensive approach to model a huge number of pedestrians and to simulate high density crowd behaviour in overcrowding places, e.g. sport, concert and pilgrimage places, and to assist engineering in the resolution of complicated problems through integrating a number of models from different research domains.

\end{abstract}

\begin{multicols}{2} 

\section{Introduction}

\lettrine[nindent=0em,lines=3]{I} n recent years, several models for the movement of crowds have been proposed.
One can distinguish between two general approaches: microscopic and macroscopic models. In the microscopic scope (range), pedestrians are treated as individual entities (particles). The particle interactions in this model are determined by physical and social rules as well as the interactions between the particles with their physical surrounding. In this context we cite,
the social-force models (see \cite{Helbing-Farkas2002} and the references therein), the cellular automata, e.g
\cite{Fukui1999, Muramatsu2000}, queuing models e.g. \cite{Yuhaski1989}, or continuum dynamic approaches like \cite{Treuille2006}. For an extensive survey of different microscopic approaches we refer to \cite{Helbing2001}. 

The origin of the microscopic models for pedestrian behaviour can be traced back to the work of Reynolds in the years 1987 and 1999 \cite{Reynolds1987,REYNOLDS1999}.  This model was extended by the sociological factors introduced in \cite{MUSSE1997}, psychological effects in \cite{PELECHANO2005}, social forces \cite{Helbing1995, CORDEIRO2005, GAYLE2009, SUD2007}, and other models of pedestrian behaviour \cite{SHAO2007, PARIS2007, YEH2008}. Many methods have been developed to describe the short range forces responsible for the collision avoidance
between nearby pedestrians. These include geometrically-based algorithms \cite{FIORINI1998, FEURTEY2000, GUY2009, SUD2008, VANDENBERG2008, VANDENBERG2009}, grid-based methods \cite{schadschneider2010stochastic, LOSCOS2003, LIN2009} and force-based methods \cite{HEIGEAS2003, LAKOBA2005, SUGIYAMA2001, SUD2007}.

In contrast to microscopic models, macroscopic models treat the pedestrians flow as continuum homogeneous mass that behaves like a fluid without considering the behaviour of single individuals. Many methods use concepts familiar from 
fluid and gas dynamics, see \cite{Henderson1971}. 
Modelling 
pedestrian dynamics as a fluid can be traced back to \cite{Fruin1971a, Fruin1971b}. More
recent models are based on optimal transportation methods \cite{Maury2009}, mean field games
(see \cite{Lasry2007} for a general introduction) or non-linear conservation laws \cite{Colombo2005}. 

\section{Pedestrian forces and motion (in PedFlow)}
PedFlow is a microscopic simulation model developed by 
L\"ohner Simulation Technologies International, Inc. (LSTI) \cite{Loehner}. In this model  individuals are treated as self-driven particles  that interact through social and physical rules. Each person in the pedestrian flow has a desired position and desired velocity and adapts his or her current velocity according to the surrounding neighbour; furthermore, each crowd member simultaneously tries to avoid collision with other crowd members and any environmental boundaries. The current velocity and direction of a single individual in the crowd is a result of the circumstances and social interactions. These social forces can be influenced by the environment, other people and internal states.

Force plays an essential role in every crowd model. It can be classified in two major categories, internal and external forces. Forces are characteristics of crowds and crowd disasters. Force has a direct effect on movement: for example force drives people out of rest to their desired velocities, or force can make deviate people from their desired direction. In dangerous crowd situations, force can be the key to decode the fatal consequences (e.g. injuries and death). Many force effects are summarized and documented by Fruin \cite{Fruin1993}.

In this approach the pedestrians are treated as "particles" moving according to Newton's law:
\begin{equation}
m_{p} \frac{d\vec{v}}{dt} = \vec{f}, \quad \frac{d\vec{r}}{dt} = \vec{v}
\end{equation}
Here $m_{p}$, $\vec{v}$ and $\vec{r}$ denote the mass, velocity and position of the pedestrian, and $\vec{f}$ the sum of all forces exerted by it or acting on it. The basic unknown that requires intensive modelling efforts in these equation is the force-vector. In the sequel, we will discuss the main forces that can be identified acting on a pedestrian.

\subsection{PedFlow force models}
In the PedFlow model, the forces that accelerate a pedestrian are the result of internal (will) and external (collision, signals, etc.) forces. An "individual alert"  tries to avoid collisions before they happen, resulting in vanishing external forces. However, if the pedestrian density increases, collision forces will appear, and the ability to move freely will be impaired. The difference between external and internal forces is too rigorous and only serves descriptive purposes in the present context.
So, what are the forces that act on the pedestrian?
\begin{itemize}
\item Internal forces (active, intentional)
\begin{itemize}
\item Will force (get there in time)
\item Pedestrian collision avoidance forces
\item Wall/Obstacle collision avoidance forces
\end{itemize}
\item External forces (passive, suffered)
\begin{itemize}
\item Physical pedestrians collision forces (physical contact)
\item Physical wall/obstacle collision forces (physical contact)
\end{itemize}
\end{itemize}

\subsection{General description of will force}
\begin{figure}[H]
\begin{center}
\includegraphics[width=\columnwidth]{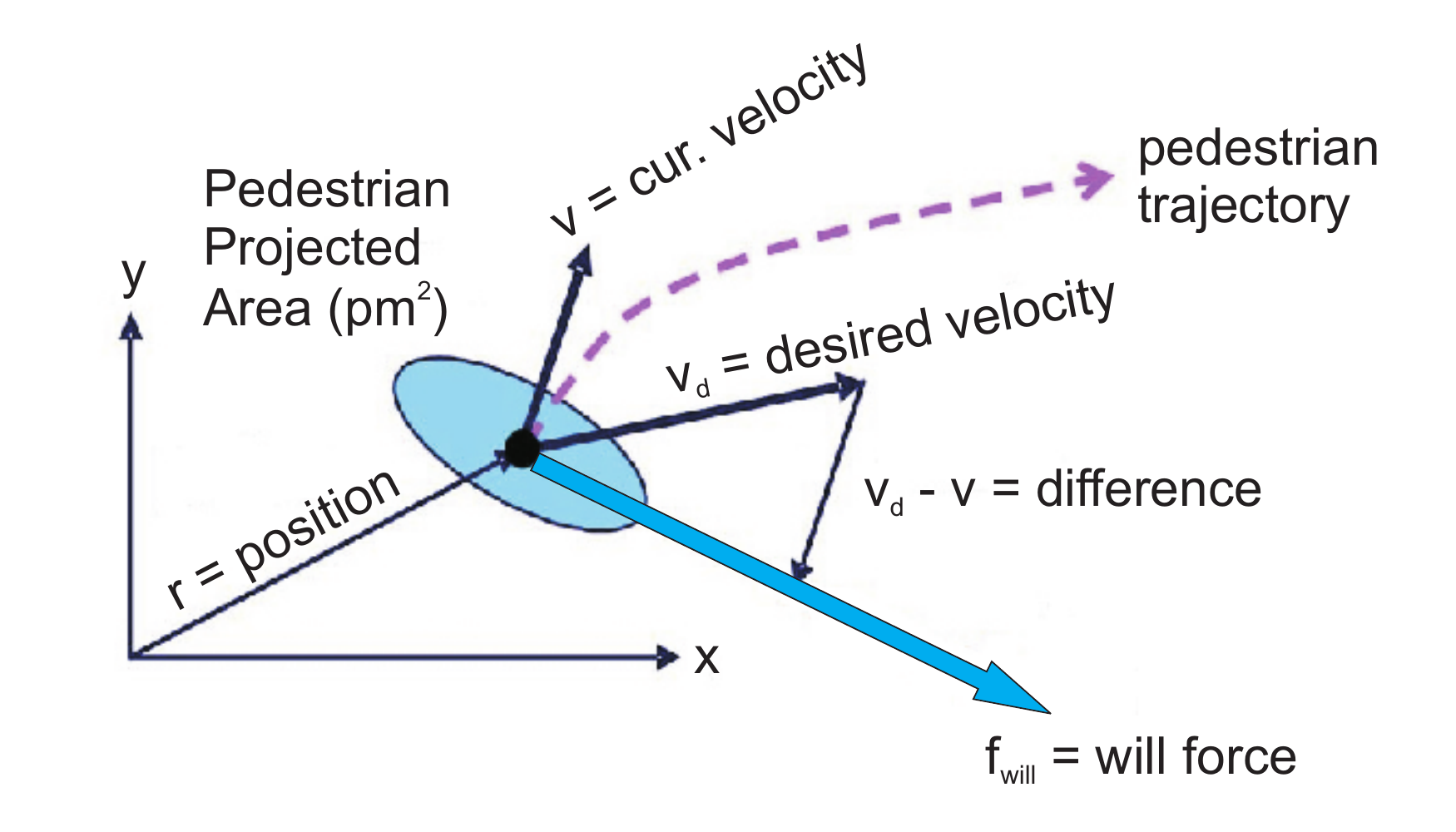}
\end{center}
\caption{The will force $\vec{f}_{will}$ denotes the self-driven force that drives an
individual towards their desired velocity.}
\label{fig:willforce-1}
\end{figure}

Each particle is programmed to have a final goal that it wants to reach. The motivation to move towards this goal is given by the driving force, or so called 
will force $\vec{f}_{will}$
and denotes the self-driven force that drives an
individual towards their desired velocity (see fig. \ref{fig:willforce-1}). The self-driven
force is based on a simple error correction term
consisting of the difference between an individuals
desired velocity, and a
desired direction $\vec{z}$ , and their actual velocity. This difference in velocity is corrected over a
specified time interval 
$\tau$, referred to as the
relaxation time, that corresponds to the finite amount of time that is required for people to react and physically
change their velocity.

In the following, we denote by will force the force that will accelerate (or decelerate) a pedestrian to achieve the velocity it desires. Given a desired velocity $\vec{v}_{d}$ and the current velocity $\vec{v}$, this force will be of the form

\begin{equation}
\vec{f}_{will} = g_{w} (\vec{v}_{d} - \vec{v} ),     
\end{equation}
where $g_{w} = f(\tau)$ is a function of "list of circumstances"

\begin{figure}[H]
\begin{center}
\includegraphics[width=\columnwidth]{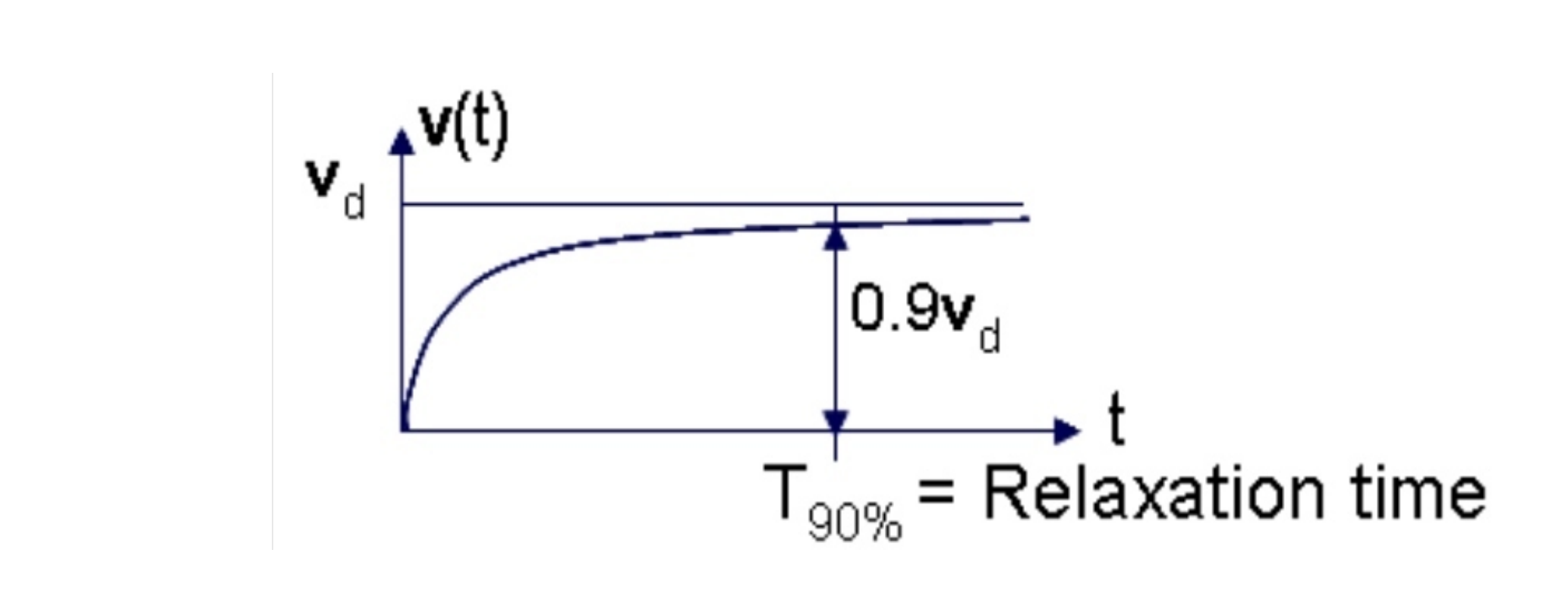}
\end{center}
\caption{Speed-up of an isolated particle to the desired velocity. This difference in velocity is corrected over a
specified time interval $\tau$
, which is referred to as the
relaxation time, that corresponds to the finite amount of time that is required for people to react and physically change their velocity.}
\label{fig:relaxationtime-1}
\end{figure}

The modeling aspect is included in the function $g_{w}$, which in the non-linear case may itself be a function of $\vec{v}_{d} - \vec{v}$. Suppose $g_{w}$ is constant, and that only the will force is acting. Furthermore, consider a pedestrian at rest. 
The pedestrian could start from rest $\vec{v}_{0} = 0$ and wants to reach the desired velocity $\vec{v}_{d}$. In this case, we have: $\vec{f}_{will} = g_{w} ( \vec{v}_{d} - \vec{v} )$, with $\vec{v}(0) = 0$ we have $m_{p} \frac{d\vec{v}}{dt} = \vec{f}_{will} = g_{w} ( \vec{v}_{d} - \vec{v} )$. The solution of this equation is given by $\vec{v}(t) = \vec{v}_{d} (1 - e^{-\alpha t})$ with $\frac{d\vec{v}}{dt}(0) = \vec{v}_{d} \alpha = \vec{v}_{d} / \tau$ and
$\alpha = g_{w} / m_{p} = 1 / \tau$
We can thus obtain $g_{w}$ by measuring the time required to reach a percentage (e.g. 90 percent) of the desired velocity, starting from rest (see fig. \ref{fig:relaxationtime-1}). This time is typically in the range of 0.5 - 1.0 s, but obviously depends on the current state of fitness, stress, climate and terrain condition, and desire to reach a goal.
We can define the function $g_{w} = m_{p} /\tau$ via "relaxation time" $0.5 < \tau < 1.0$ s.
Typical values for the desired velocity $\vec{v}_{d}$ and relaxation time $\tau$ are
	   $v_{d}$ = 1.35 m/s and 
	     $\tau$ = O(0.5 to 1.0 s). 
Medicinal facts and human experience show that older pedestrians walk more slowly than
younger pedestrians. This effect has a significant impact on the "relaxation time" of a single pedestrian, hence it is easier to assess that younger pedestrians reach their desired velocity faster than older pedestrians. 

\subsection{Internal forces}
\subsubsection{Internal forces \textendash  far range}
\begin{figure}[H]
\begin{center}
\includegraphics[width=\columnwidth]{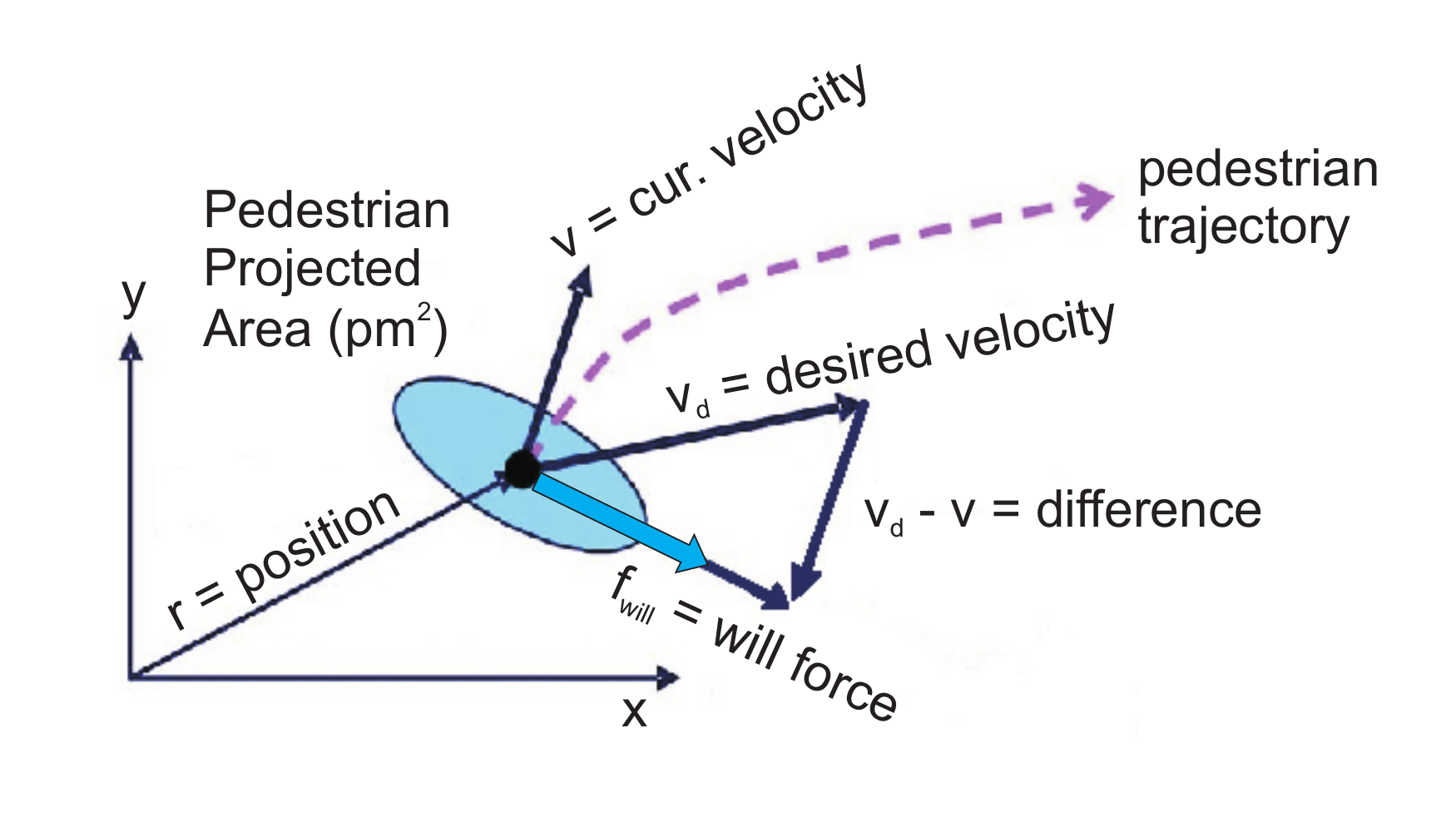}
\end{center}
\caption{Desired velocity. In the real world, each pedestrian has a desired velocity $\vec{v}_{d}$ towards a desired destination and adapts the current velocity according to the environmental situation.}
\label{fig:desiredvelocity-1}
\end{figure}

In normal cases, the will force $\vec{f}_{will}$ takes the direction of the desired velocity $\vec{v}_{d}$ as shown in figure  \ref{fig:desiredvelocity-1}.

\begin{figure}[H]
\begin{center}
\includegraphics[width=\columnwidth]{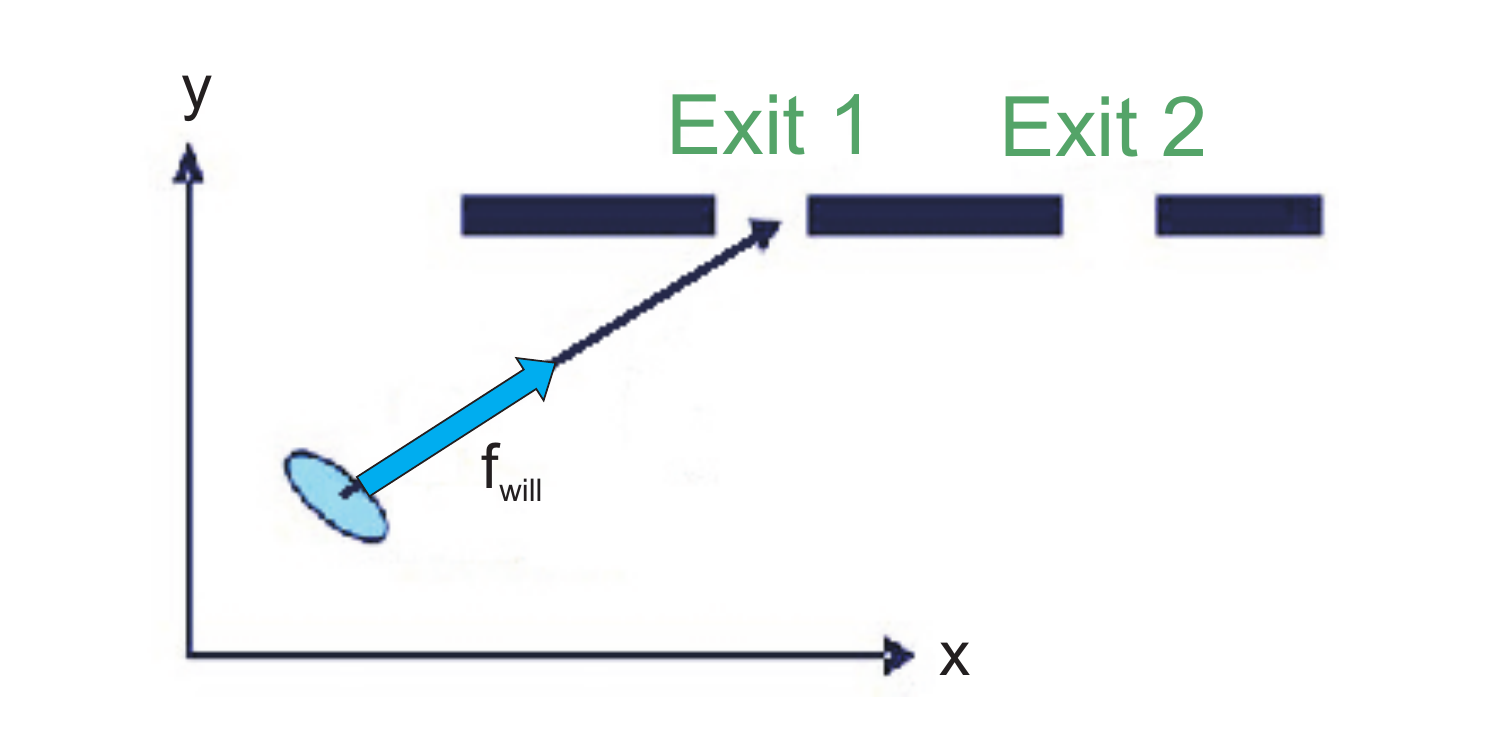}
\end{center}
\caption{Go to nearest exit.}
\label{fig:nearestexit-1}
\end{figure}

\begin{figure}[H]
\begin{center}
\includegraphics[width=\columnwidth]{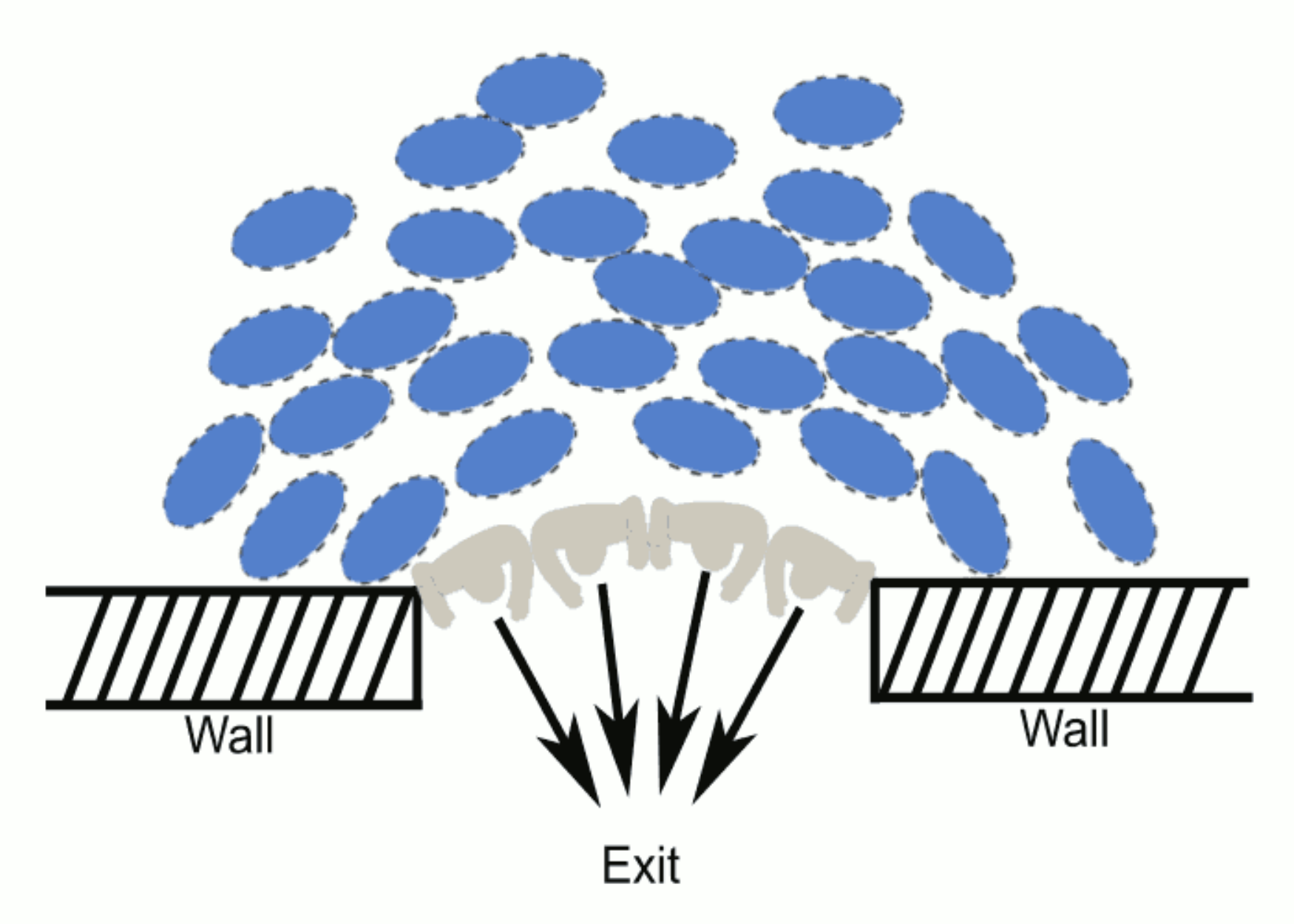}
\end{center}
\caption{Scheme of an arc formation at openings and bottleneck areas. The full ellipses represent pedestrians moving in the same directions. The pedestrians block one another in the direction of the movement and congest the passage area.}
\label{fig:arcformation-1}
\end{figure}

Traditionally the will forces are specified by a desired velocity of the pedestrian and the trajectory.
The will force will be implemented as a vector on the simulation model. However, in the motivation to reach certain places, there are tasks in which the desired motions of individuals are specified by the state of system (list of  circumstances) rather than time. For such tasks, desired velocity field control has been proposed. In this chapter, a method to extend our information about the forces that drive the individuals from rest to their desired velocities is evaluated. 
\begin{itemize}
\item In the case of group movement, the will force takes the direction of the group leader.

\begin{figure}[H]
\begin{center}
\includegraphics[width=\columnwidth]{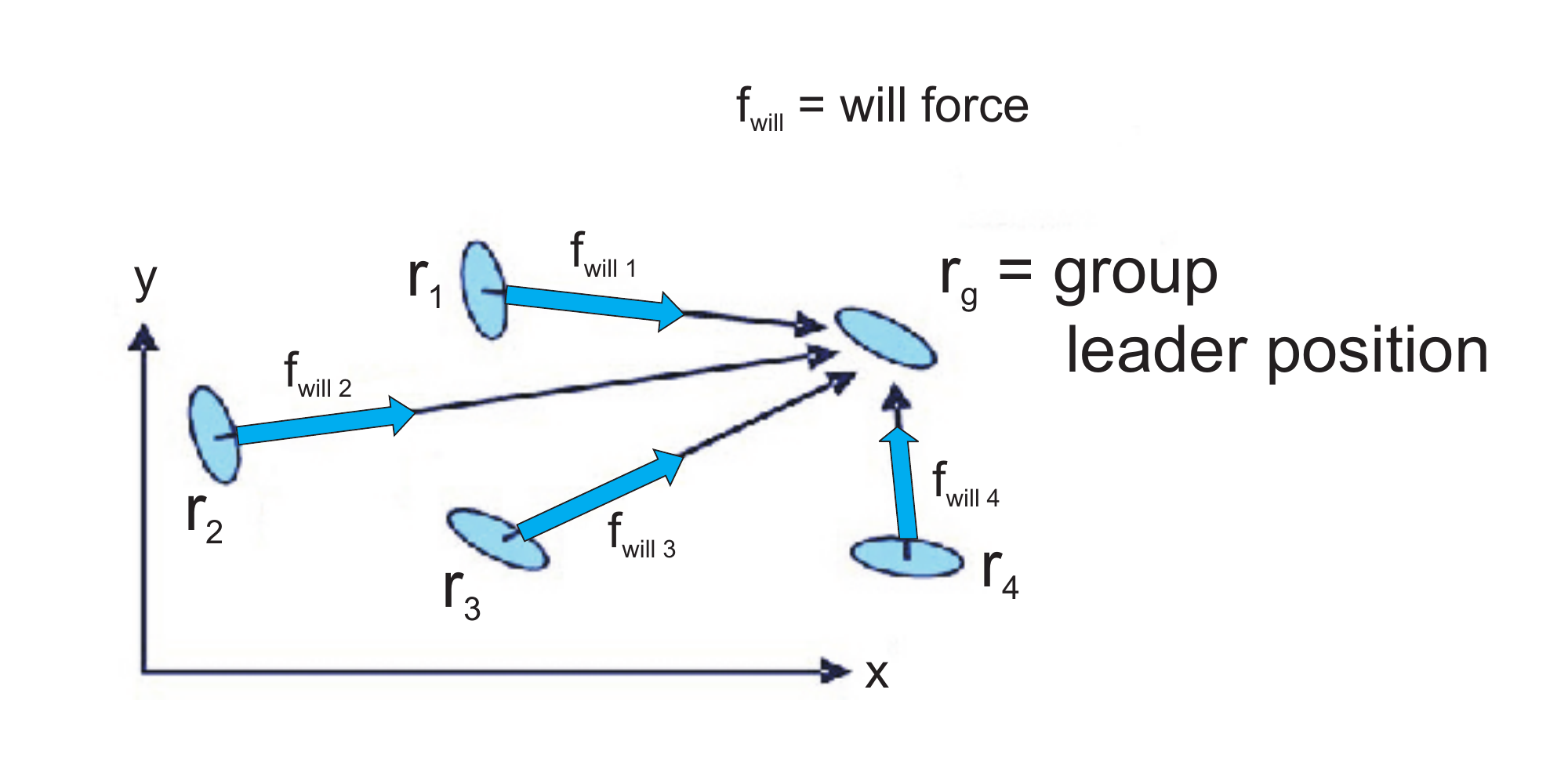}
\end{center}
\caption{"Follow me".}
\label{fig:followme-1}
\end{figure}
Figure \ref{fig:followme-1} illustrates a group of pedestrians following their leader. A "group" is defined as a 
physical collection of people following the same route, but who may or may not be part of the same social group, and 
a "subgroup" is defined as people within the same physical "group" who want to stay together, like friends or family members. Several studies have revealed that smaller subgroups constitute the majority of the people in a 
crowd. But very few studies are able to model the "subgroup" behaviour. A particular subgroup 
concerns pedestrians who hold hands. It is recognized that holding hands is the most effective way of keeping children safe
from traffic injury \cite{Pecol2013}. All the group members take the desired position and the desired velocity of the group leader.
\item In the case of an emergency or dangerous situation, the will force takes the direction of the nearest exit and this has dramatic consequences on the behaviour of pedestrians at the exit. They block each other and this builds up a rigid arc from which no one can escape (see fig. \ref{fig:nearestexit-1} and \ref{fig:arcformation-1}).
\end{itemize}
Figure \ref{fig:nearestexit-1} demonstrate the situation of pedestrians in case of emergency or dangerous situation.
The will force points to the direction of the nearest exit. This picture illustrates the situation when people take the proper precaution to prevent potential hazards when they are alerted. For example, in case of an emergency, evacuating the building immediately by using the closest exit and going to the designated meeting location.
Panic breaks out, if many individuals  attempt to escape an emergency area at the same time, often the emergency exits are blocked in a short time, and the situations have a fatal consequence for survival and leads to injuries or death through crushing people in a crowd. Real world example (fire safety: General Evacuation Procedures): Learn the location of the two nearest exits from your work area (often these are down staircases)\cite{Oxford2013}.

The priority in the emergency or dangerous situations like fire in a building or earthquake are designed to reach each particle the nearest exit in a short time. In first step the particle choose their priority destination after that they can find the beeline towards the nearest exit.

\subsubsection{Internal forces \textendash  intermediate range}
\begin{figure}[H]
\begin{center}
\includegraphics[width=\columnwidth]{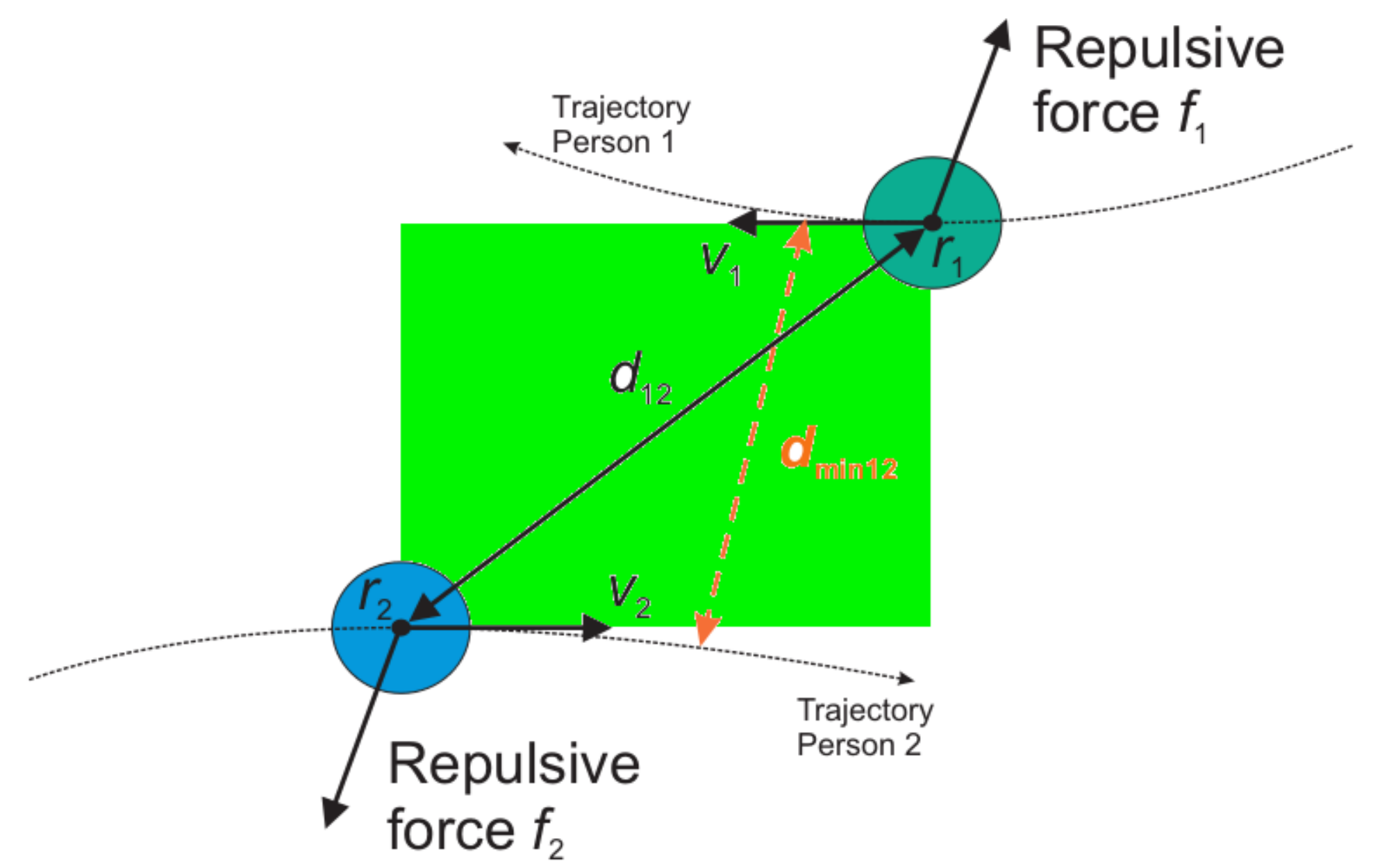}
\end{center}
\caption{Collision avoidance rectangle of influence. Illustration of pedestrian-pedestrian collisions avoidance forces. This force is referred as repulsive force and reaches its max. value at the shortest distance between the particles.}
\label{fig:CollisionAvoidance-2}
\end{figure}

(a) Collision avoidance forces\\
In the collision avoidance process, autonomous particles need to discover the environment to
avoid static and dynamic obstacles.
At any time the distance between 
each obstacle, wall, and the particle $i$ are computed, go both of them closer to each other, then
calculate the angle between particle $i$'s desired direction and
the line joining the center of particle $i$ and the obstacle. This effect can be seen within the
rectangle of influence (figure \ref{fig:CollisionAvoidance-2}). The distance and the
angle play a significant role to assess how relevant
the obstacle is to the trajectory. 
When travelling the environment, particles also update their perceived crowd density, which is necessary for their decision making process.

We denote by intermediate range forces that change the motion of an individual in order to avoid an encounter with another. The observation employed here is that all of us will try to avoid a collision by moving away from an encounter long before it happens. The nature of these forces is such that they act mainly in a direction normal to the current motion of the individual, and therefore do not lead to a significant decrease in velocity. Moreover, this force is only active if a collision is sensed.  

Using as a starting point the situation depicted in figure \ref{fig:CollisionAvoidance-2}, given two pedestrians with current coordinates and velocities $\vec{r}_{1}$, $\vec{v}_{1}$ and $\vec{r}_{2}$, $\vec{v}_{2}$, an encounter may be computed by evaluating the time increment $\Delta t$ at which the distance between the two is minimized, i.e., $[ \vec{r}_{1} + \Delta t \vec{v}_{1} - ( \vec{r}_{2} + \Delta t \vec{v}_{2} )]^{2}   \rightarrow  min$. This results in
$[ ( \vec{r}_{1} - \vec{r}_{2})  + \Delta t_{m} ( \vec{v}_{1} - \vec{v}_{2} ) ]•(\vec{v}_{1} - \vec{v}_{2}) = 0 , 
$
or $\Delta t_{m} = - \frac{(\vec{r}_{1} - \vec{r}_{2})•( \vec{v}_{1} - \vec{v}_{2})}{(\vec{v}_{1} - \vec{v}_{2})•( \vec{v}_{1} - \vec{v}_{2})}
$. The minimum distance is given by
$\delta_{min} = \vert (\vec{r}_{1} - \vec{r}_{2}) + \Delta t_{m} (\vec{v}_{1} - \vec{v}_{2}) \vert$ . 
Obviously, the force will only become active if  $\Delta t>0$. The general nature of this force is such that it will decrease with distance. How exactly this decrease function looks like is unknown at the present time (it may even be random). We have used a function based on the normalized distance between the pedestrians $
d  = \frac{\vert (\vec{r}_{1} - \vec{r}_{2}) \vert}{R_{1}}$ , where $R_{1}$ is the characteristic radius of pedestrian 1, which is of the form: $f = f_{max} \frac{1}{( 1 + d^{2})}$. 
This simple repulsion force may be refined further by considering the directions along and normal to the current velocity vector $\vec{v}_{1}$, denoted by $\hat{e}_{t}$ and $\hat{e}_{n}$, respectively. We can then split the normalized distance $d$ into a tangential and normal component:
$d_{t}  = \frac{\vert \hat{e}_{t}. (\vec{r}_{1} - \vec{r}_{2}) \vert}{R_{1}}$,
$d_{n}  = \frac{\vert \hat{e}_{n}. (\vec{r}_{1} - \vec{r}_{2}) \vert}{R_{1}}$.
This leads to tangential and normal forces of the form: $\vec{f}  = - f_{max} \frac{1}{( 1 + d_{t}^{2} )} \hat{e}_{t} - f_{max} \frac{1}{( 1 + d_{n}^{2} )} \hat{e}_{n}$. 
The value of $f_{max}$ can be related to the relaxation time, i.e. it is not too difficult to obtain. We have used  $f_{max} /m_{p}$ = O(4) $\approx$ 4 [m/sec$^{2}$].

\begin{figure}[H]
\begin{center}
\includegraphics[width=\columnwidth]{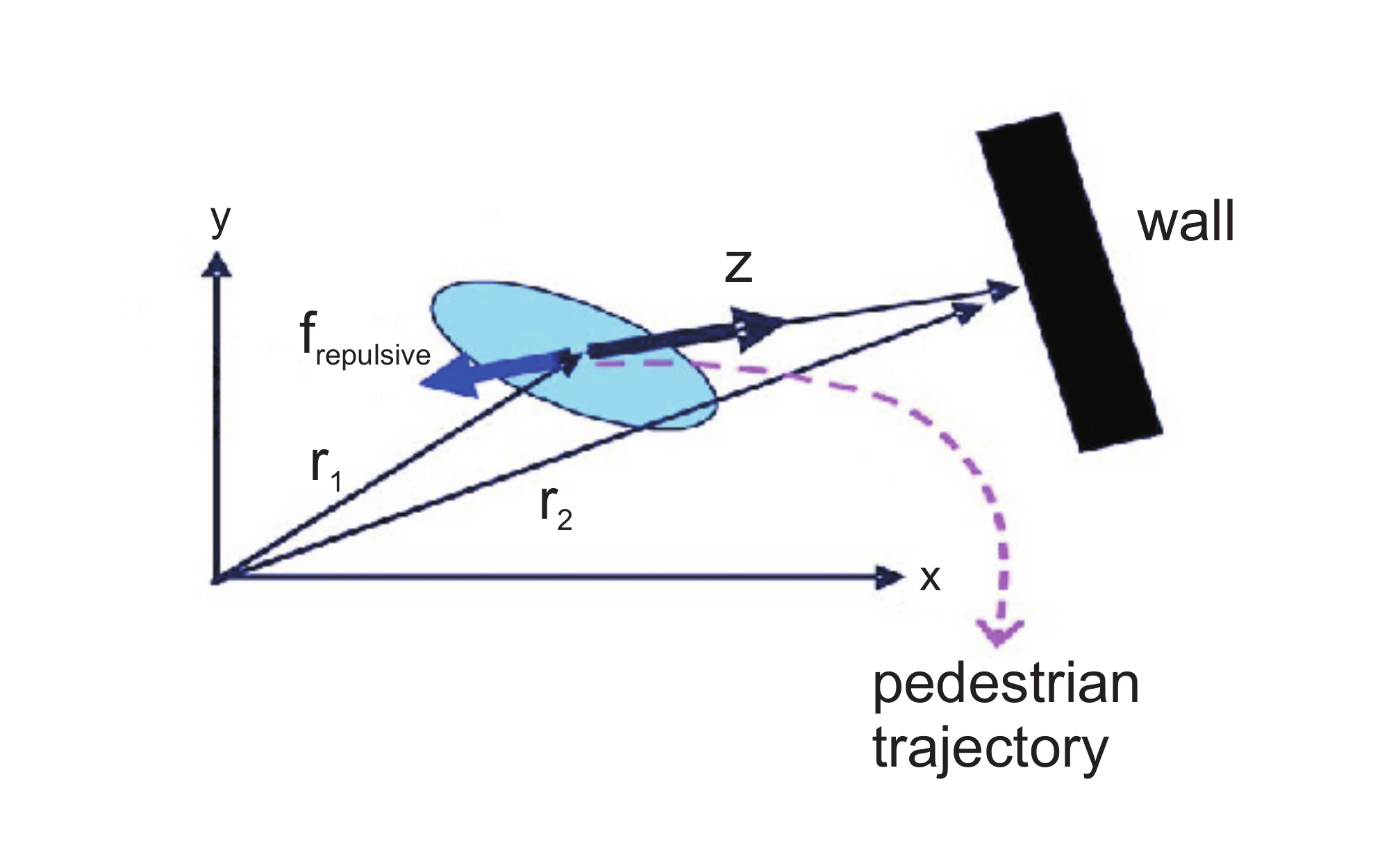}
\end{center}
\caption{Avoiding a wall. This picture illustrate the pedestrian-obstacle collision avoidance. The blue solid arrow shows the direction of the repulsive acting force $f_{repulsive}$ and the dashed arrow indicates the trajectory path of the particle.}
\label{fig:wallavoiding-1}
\end{figure}

(b) Avoiding a wall\\
Avoidance forces $f_{repulsive}$ are calculated only for relevant obstacles,
walls and agents: i.e. those falling within the rectangle of
influence.
The avoidance force for obstacle W is illustrate in figure \ref{fig:wallavoiding-1}. The calculation is done in the same way as for the force of the pedestrian-pedestrian repulsion and the same formulas are used. Only the pedestrian-pedestrian distance has to be replaced by the distance of the pedestrian from the wall. 

\subsection{External forces}
\subsubsection{External forces \textendash  near range}
\begin{itemize}
\item Pedestrian collision: 
For more realistic crowd simulation and for a description of realistic counter-flows and overtaking behaviours,
many attributes influencing the behaviour of people in the real world will be considered. This approach has the ability to simulate human
behaviour by setting parameters obtained from many observations related to real human
movement. In avoiding an obstacle the factors affecting the tangential forces are:
\begin{itemize}
\item The distance between the particle and any obstacles.
\item The relative direction and movement of the agents i.e. the direction between the desired
velocities vectors ($\vec{v}_{1}$ and $\vec{v}_{2}$ for example).
\item Crowd density:
If an agent enter the interaction rectangle, a tangential force (see fig. \ref{fig:CollisionAvoidance-2}) will be active to change the direction of movement and make a curve in the path to avoid  collision. The angle between two agents is used to simulate the individual decision making how to react to a previous collision. For example, in a walking floor, if there is enough space between us and another person, none of us would change the direction. In case of many persons walking in the corridor, the majority of people tend to move to the right. Therefore, when the velocity vectors are almost collinear, the tangential forces pointing to the right.
In case an agent $ 1 $ detects agent $ 2 $ and agent as possible obstacles, then the system computes the distance vector
towards agent $1$ as follow ($d_{21}$). Agent $2$ is
farther away from $1$, but since it is moving against agent $1$,
the system gives this obstacle higher priority.
\end{itemize}
\begin{figure}[H]
\begin{center}
\includegraphics[width=\columnwidth]{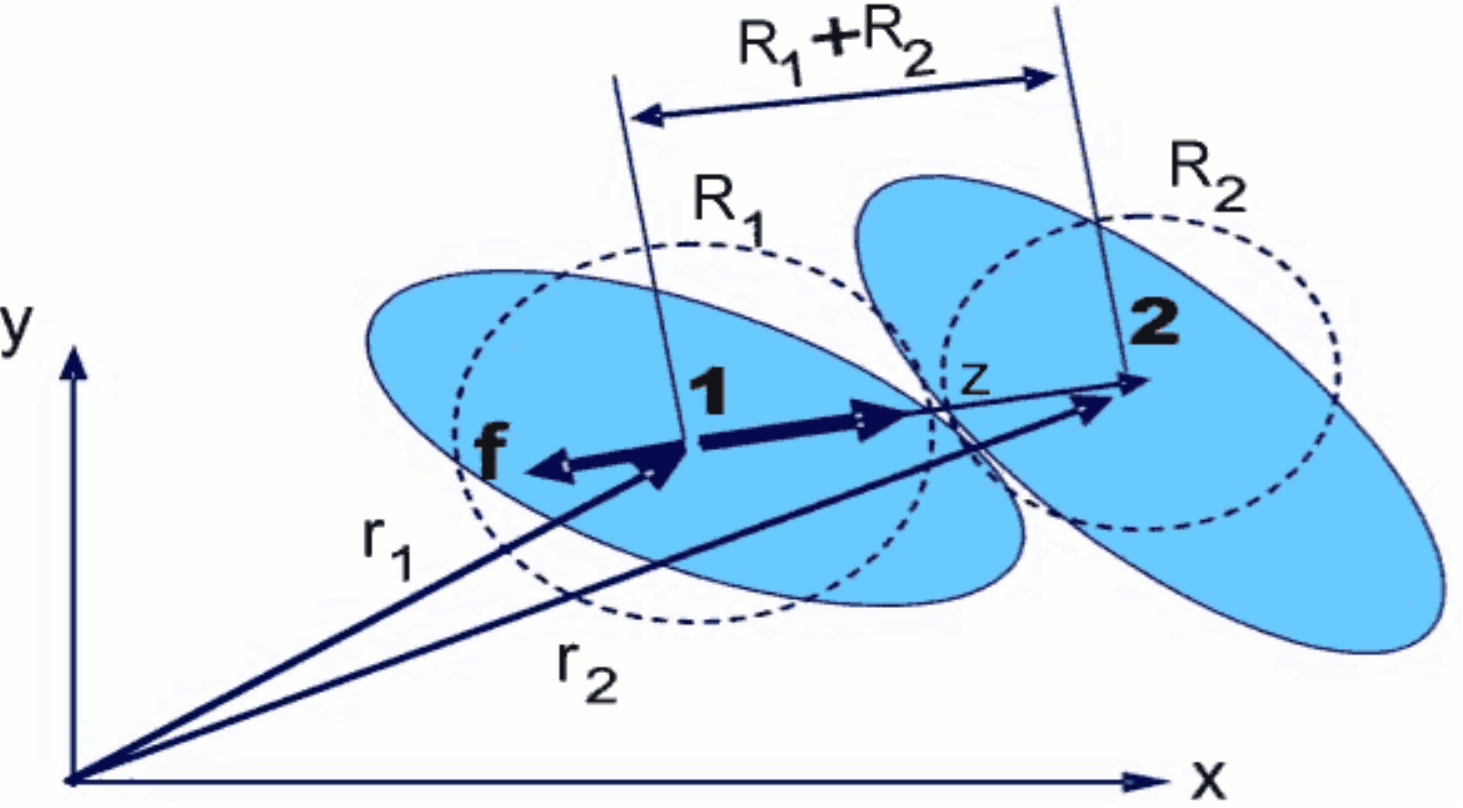}
\end{center}
\caption{Pedestrian collision. Schema illustrate pedestrian-pedestrian body contact or collisions forces. This situation indicate a body friction or body deformation.}
\label{fig:pedestriancollision-1}
\end{figure}

Once a pedestrian has moved close enough to obstacles or other pedestrians, his velocity will be decreased markedly, (see fig. \ref{fig:pedestriancollision-1}). Unlike the long-range forces, these forces act in the direction of the normalised difference vector  
$\vec{z} = \frac{(\vec{r}_{2} - \vec{r}_{1})}{\vert \vec{r}_{2} - \vec{r}_{1} \vert}$.
We have used, as before, a force of the form
$\vec{f}  = - f_{max} \frac{1}{( 1 + d^{2})} \vec{z}$, 
although the exact nature of this force is unknown at the present time. As before, the value of $f_{max}$ can be related to the relaxation time. We have used  $f_{max} /m_{p}$ = O(4) $\approx$ 4 [m/sec$^{2}$].

\item Collision with a wall or obstacle: 
A stream of pedestrians will be enclosed by the walls in or around the structures in which it occurs (see fig. \ref{fig:repulsivecontactforce-1}). We have assumed that this force acts in the direction of the gradient of the distance-to-wall function $d_{w}(x)$. This function, which measures the closest distance to a wall from any given location $r$, is assumed to be known. In practice, it is evaluated in a pre-processing step. As before, we have used, for lack of any deeper knowledge, a function of the form:
$\vec{f}_{w} = - f_{max} \frac{1}{( 1 + (d_{w} / R)^{2} )}. \nabla d_{w}$,	
with $f_{max}/m_{p}$ = O(4) $\approx$ 4 [m/sec$^{2}$].
\begin{figure}[H]
\begin{center}
\includegraphics[scale=0.50]{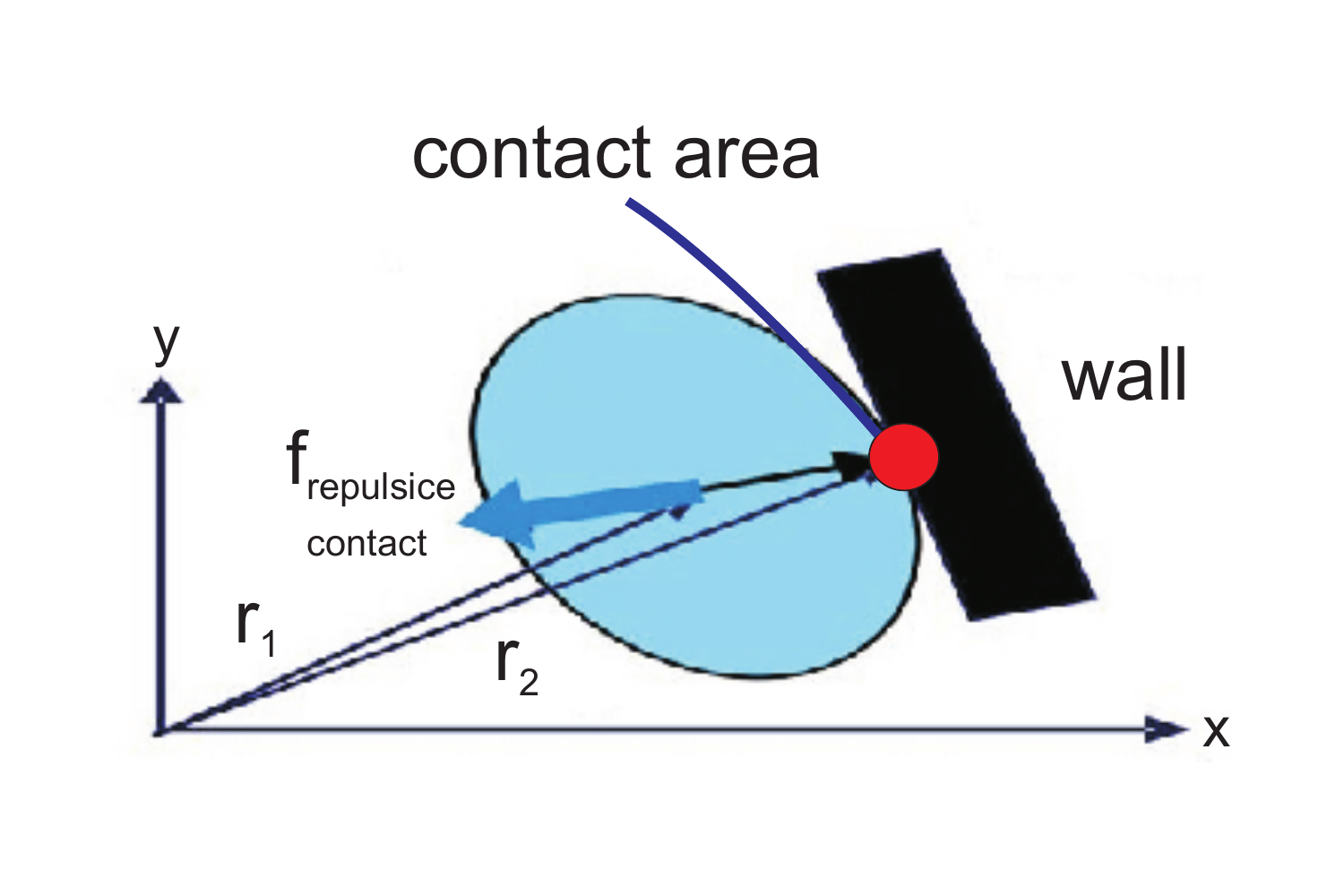}
\end{center}
\caption{Repulsive contact force. The solid blue arrow indicates the direction of the contact repulsive force.}
\label{fig:repulsivecontactforce-1}
\end{figure}

$\vec{f}$ denotes the individual-individual
interaction force or individual-obstacle force,
$f_{w}$ denotes the collision force due to the 
wall (out of a total of $M$ wall surfaces).
\end{itemize}
\subsection{Data structures}
\subsubsection{Data carried by the individual}
The data concerning a single individual are classified in three categories: physical, emotional and ethnic data. This data, which includes items such as the pedestrian's height, width, fitness, nationality, current physical and mental state, as well as the current motion data (location, velocity, etc.) is stored in arrays attached directly to the pedestrians. In this way, the data becomes easily accessible at any time. In order to save storage, the individual's height, width and fitness are taken from a data-base, i.e. through a table look-up. In this table, the data belonging to a limited number of representative groups is stored, and the individual is tagged as belonging to one of these groups. In order to enhance realism, a random variation number for personal data is also attached to each individual. The savings in storage accrued by this indirect storage assignment are considerable: if we have Nd personal data items, conventional storage would require O(NdNp) items, whereas indirect data assignment only requires O(2Np+NdNg) items. This implies that even for moderate realism, with O(10) data items per person, indirect storage provides savings of O(5).
All this data is necessary for more realism in the pedestrian motion simulation. 
\subsubsection{Geographic data}
The environment where the agents navigate is represented in the system as geographic data. The geographic data contain information about the platform within which the particles move, items such as terrain condition (inclination, 
corridors, escalators, obstacles, etc.), climate data (temperature, humidity, rain, visibility), signs, as well as doors, entrances and emergency exits. All this data is stored in a so-called background grid consisting of triangular elements. This background grid is used to define the geometry of the problem, and is generated automatically using the advancing front method \cite{Loehner2008}. All geometrical and climatological data is attached to this grid. 

\subsubsection{Fixed triangular background grid}
\begin{figure}[H]
\begin{center}
\includegraphics[width=\columnwidth]{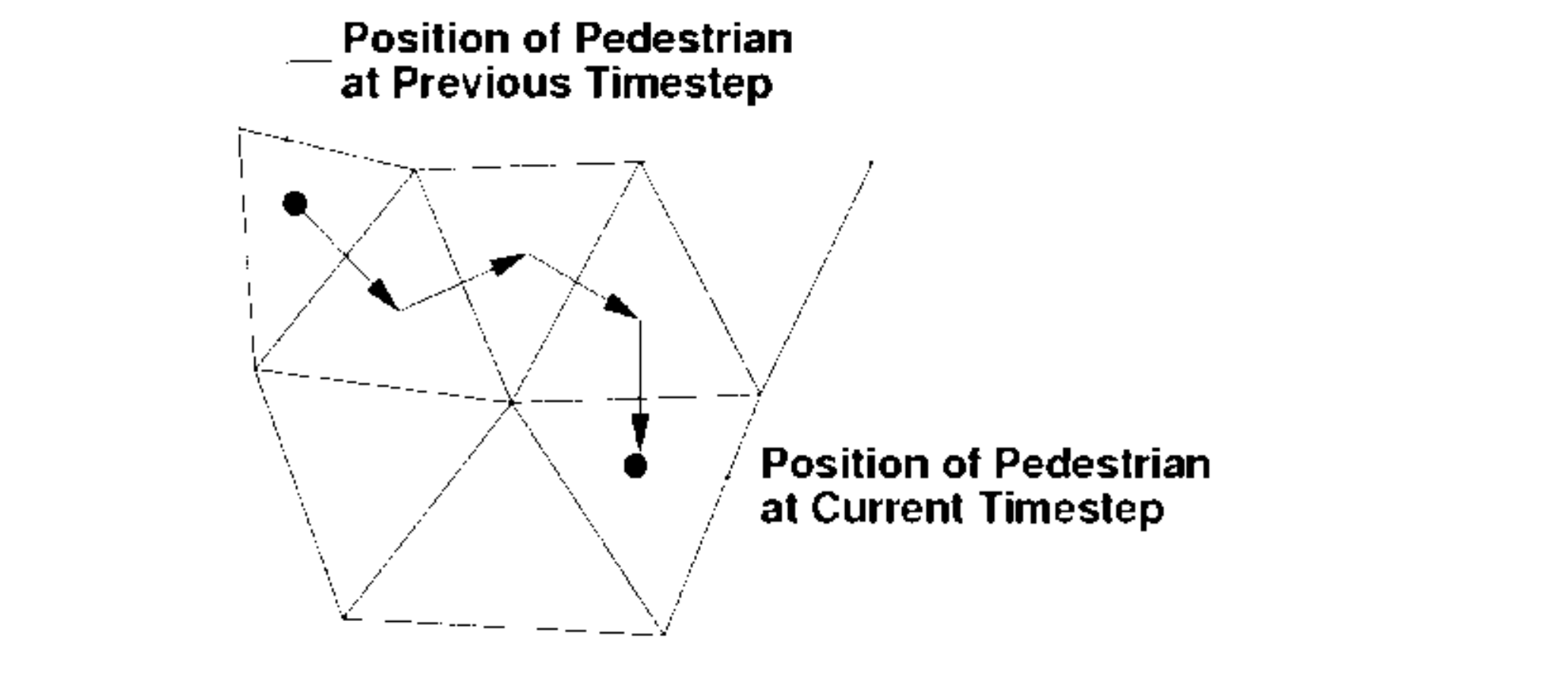}
\end{center}
\caption{Background Grid. All information about the terrain surrounding where the particle moves will be generated with a fixed triangular background grid \cite{Loehner}.}
\label{fig:backgroundgrid-1}
\end{figure}

The geography is stored at (fixed) background nodes
and used to define the geometry.
The mesh is generated automatically
and the data is interpolated (linearly) to the pedestrian position (see fig. \ref{fig:backgroundgrid-1}).

\subsection{Pedestrian data}
\begin{itemize}
\item Pedestrian "personal" data types
\begin{itemize}
\item Personal Data (height, width, fitness, nationality, familiarity with surroundings, objectives, etc.)
\item Current Physical and Mental State
\item Current Motion Data (location, velocity, etc.)
\end{itemize}
\item Pedestrian Data Storage
\begin{itemize}
\item Data table for Geno-Types and Objectives/Paths
\item Pedestrian Arrays for Actual Data
\item Pedestrian "personal" Data types
\end{itemize}
\end{itemize}

\subsubsection{Neighbour data}
\begin{figure}[H]
\begin{center}
\includegraphics[width=\columnwidth]{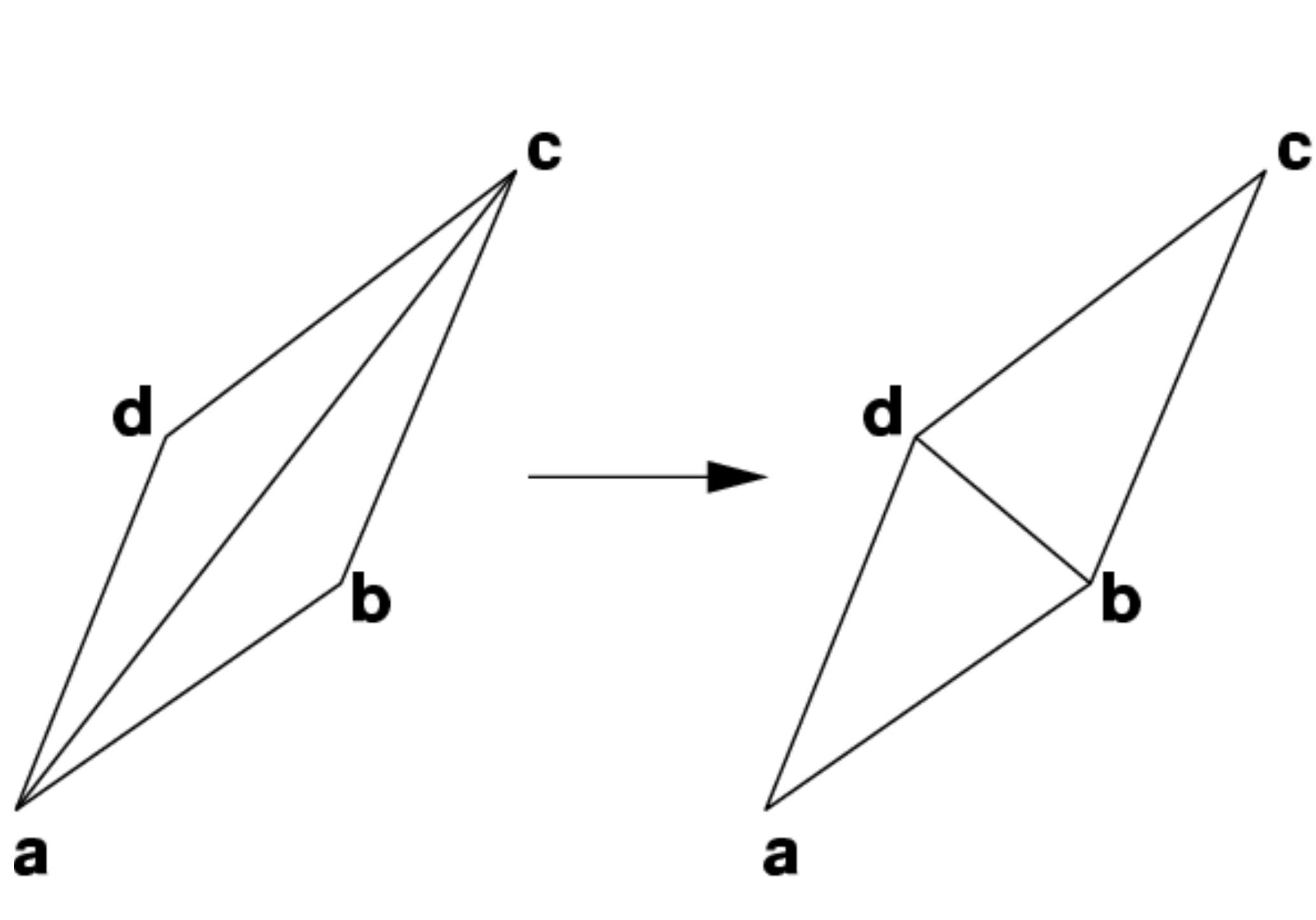}
\end{center}
\caption{Diagonal Swap Using Min-Max Criterion.}
\label{fig:diagonalswap-1}
\end{figure}

By far the most time-consuming portion of pedestrian simulations is the evaluation of the interaction between nearest neighbours. The nearest neighbours of every pedestrian must be identified and accounted for at every time step. Assuming an arbitrary cloud of points, a Voronoi tessellation, or its dual, the Delauney triangulation, uniquely defines the nearest neighbours of a point. The Delauney criterion states that the circumcircle of any triangle does not contain any other point. Another unique way to define a mesh is by minimising the maximum angle for any combination of triangles adjacent to an edge, (see fig. \ref{fig:diagonalswap-1}). 

\begin{figure}[H]
\begin{center}
\includegraphics[width=\columnwidth]{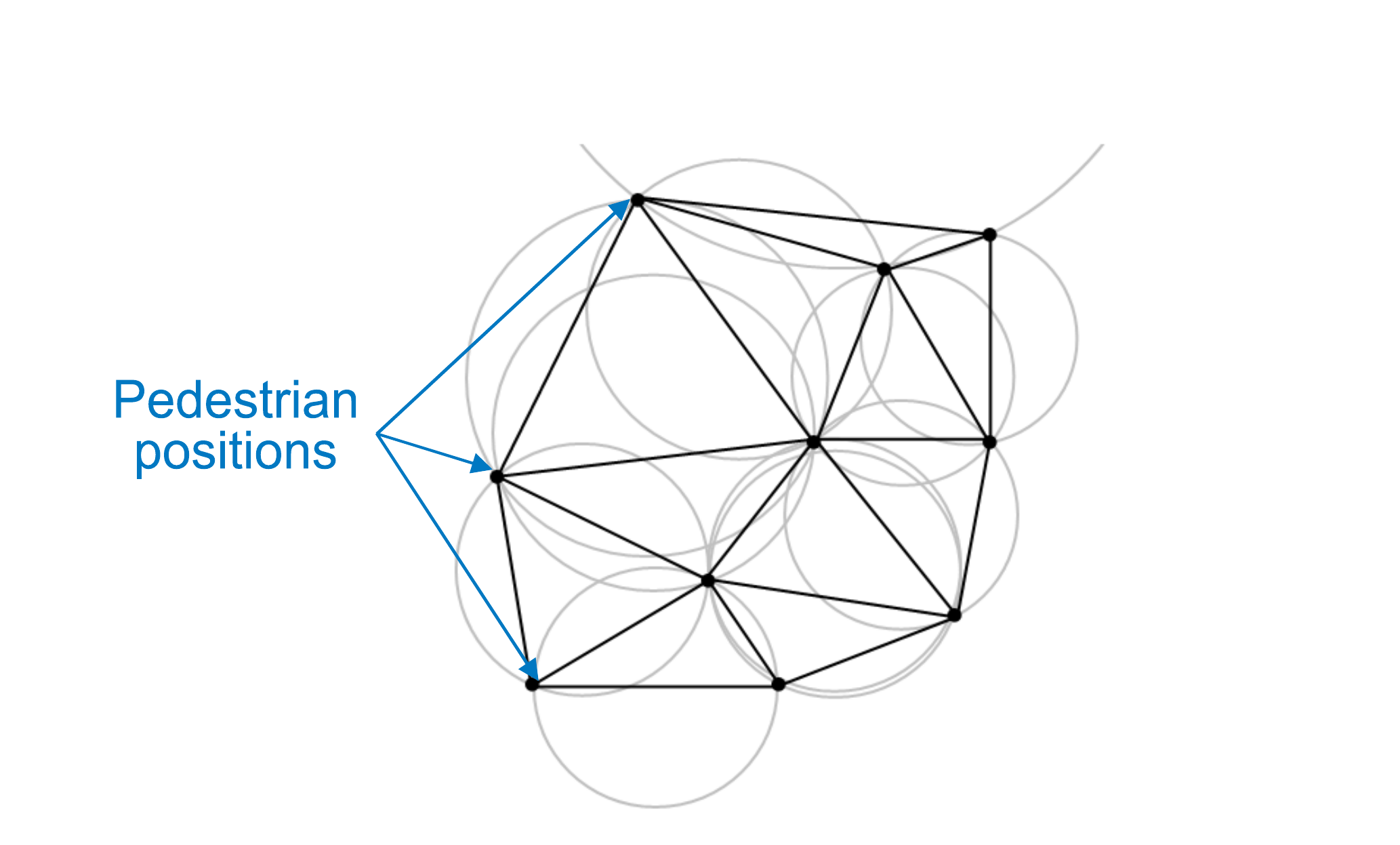}
\end{center}
\caption{Neighbour data. The points represent the particle positions and the circles indicate the particle-particle interaction ranges. Every particle can see and identify the nearest neighbours (at each $\Delta t$). In this system pedestrians are moving like a cloud of points,
Delauney triangulation (updated) \cite{Delaunay1934}. }
\label{fig:neighbordata-1}
\end{figure}

For dynamically moving points, the Delauney or min-max criterion will be violated in parts of the mesh. Every so often (e.g. after every time step), the mesh must be modified in order to restore it. For 2-D grids like the ones contemplated here, this is best achieved by flipping diagonals until the Delauney or min-max criterion have been restored. An edge-based data structure that is well suited for this purpose stores the two points of the edge, the neighbours on either side of the edge, as well as the four edges that enclose the edge, (see fig. \ref{fig:neighbordata-1}). For boundary edges some of these items will be missing, making it easy to identify them.
 
On the basis of this method the particle can identify the nearest neighbours (at each $\Delta t$).
\subsection{Example: walking through a narrow passage}
\begin{figure}[H]
\begin{center}
\includegraphics[width=\columnwidth]{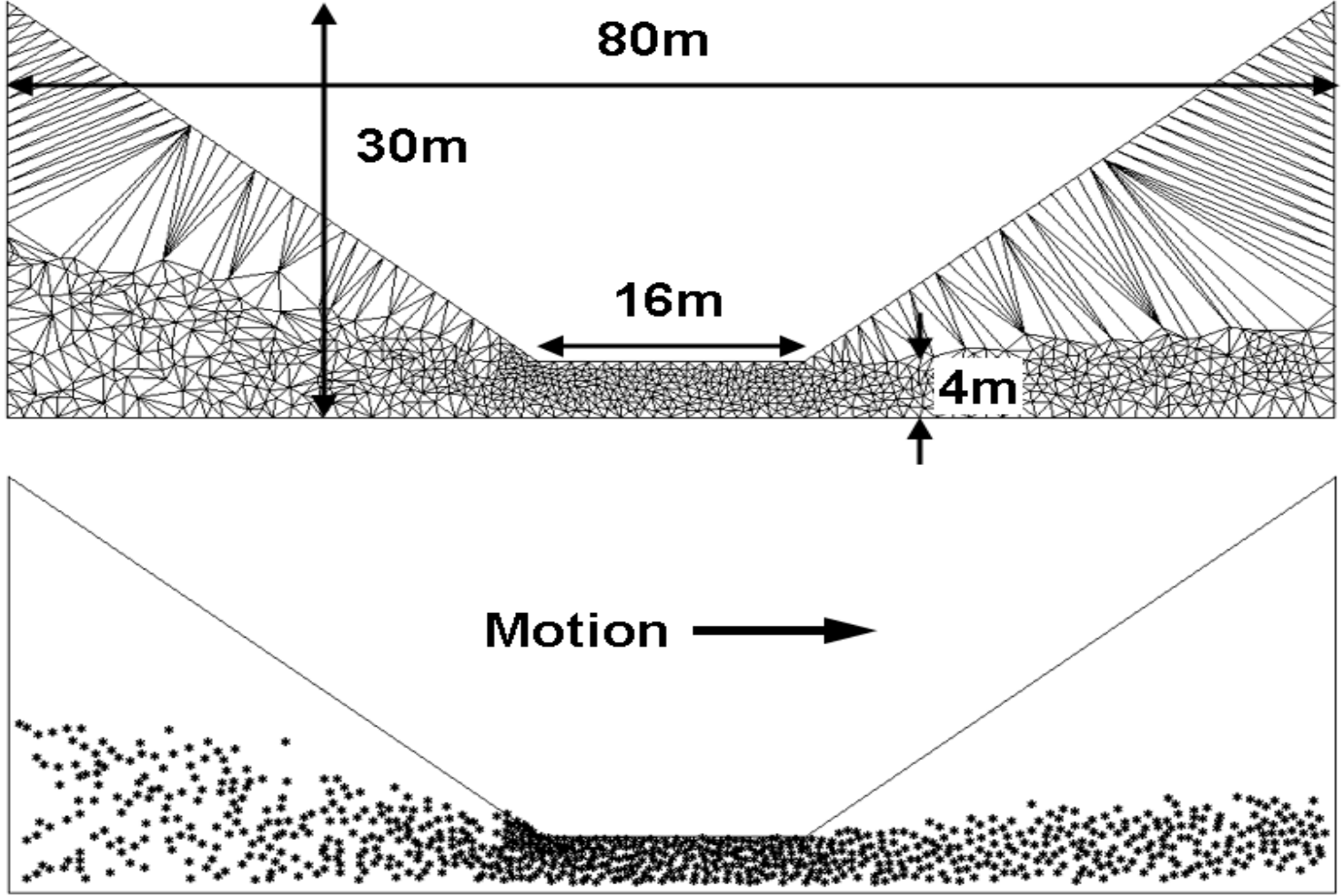}
\end{center}
\caption{Moving grid and position snapshot. This picture shows pedestrians moving through a narrow passage like a cloud of points,
Delauney triangulation (updated) \cite{Delaunay1934}. The fine triangulation indicates high density of pedestrians.}
\label{fig:movinggrid-1}
\end{figure}

This background grid is used to define the geometry of the problem, and is generated automatically using the advancing front method \cite{Loehner2008, Loehner1987, Loehner1990}. All geometrical and climatological data are attached to this grid. This implies that the amount of data stored and used for pedestrian movement depends only on the level of details stored in this mesh, and is proportional to the number of elements in it. For obvious reasons, the size of this mesh (number of elements and points) should be limited to the necessary and available amount of information required for the simulation. One can see that the mesh will be very fine in areas where the density of pedestrians gets higher (see fig. \ref{fig:movinggrid-1}).
At every instance, a pedestrian will be located in one of the elements of the background grid. This 'host element' is updated continuously with the nearest neighbour tracking procedure. 

Pedestrian motion in passages is one of the few cases where reliable empirical data exists. In order to assess the validity of the proposed pedestrian motion model, a typical `passage flow was selected. The geometry of the problem is shown in figure\ref{fig:movinggrid-1}. Pedestrians enter the domain from the left and exit to the right. For this case, each pedestrian has the goal of reaching first the entrance of the passage, then to traverse it to the other end, and finally to exit on the right. A typical snapshot during one of the simulations is shown in figure\ref{fig:movinggrid-1}.
 
\begin{figure}[H]
\begin{center}
\includegraphics[width=\columnwidth]{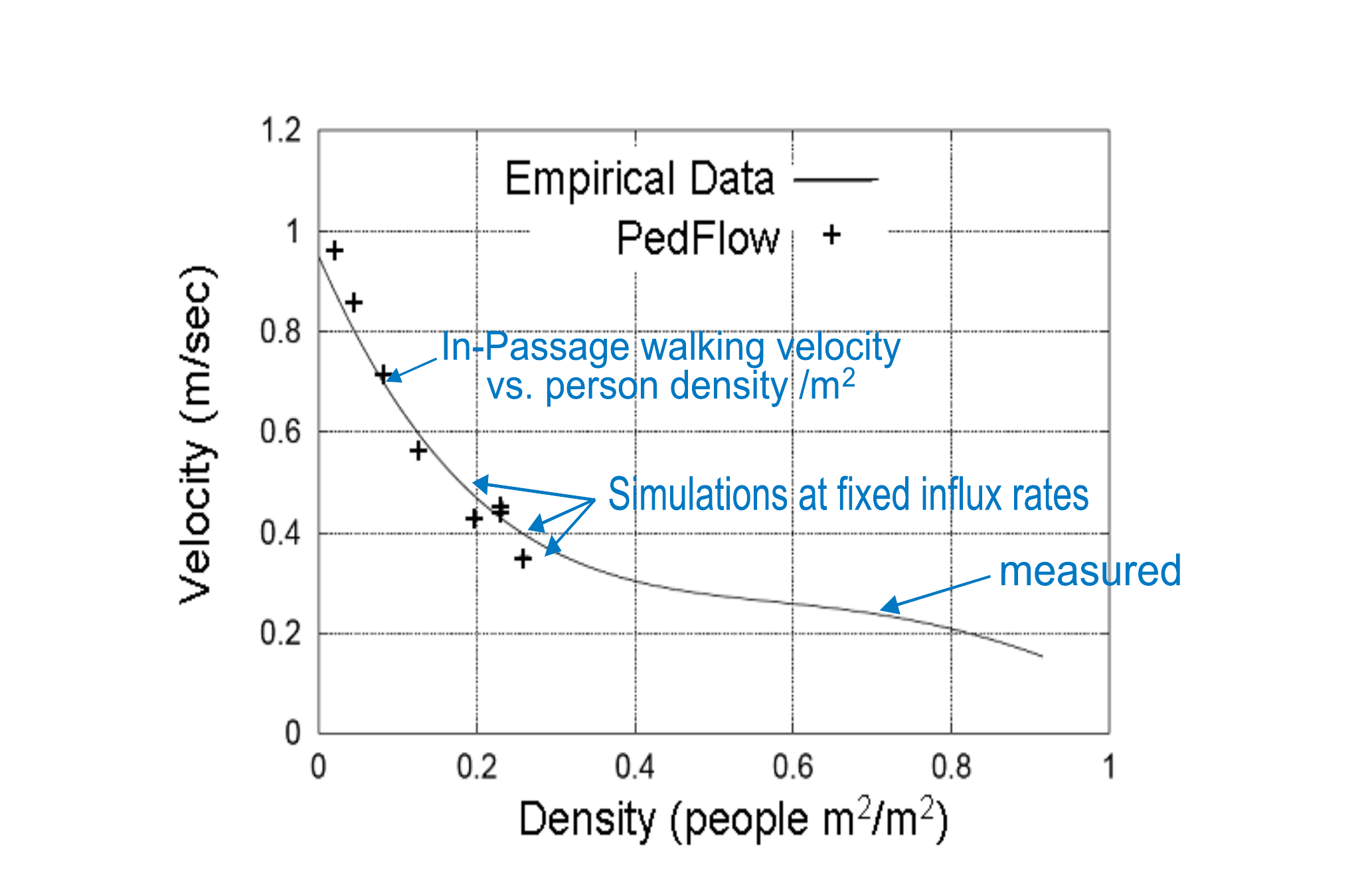}
\end{center}
\caption{Velocity-density fundamental diagram: Decrease of the pedestrian local velocity  $\vec{v} (\vec{r}, t)$ in the passage area as a function of the local density $\rho(\vec{r}, t)$. The measured data are represented by crosses, they are consistent with the empirical data of Predtetschenski and Milinski  \cite{Predtetschenski-Milinski1969} represented by the solid line.}
\label{fig:velocitydensity-11}
\end{figure}

The pedestrian parameters chosen were as follows:
\begin{itemize}
\item Desired velocity $v_{d}$ = 1 $\pm$ 0.02  m/s;
\item Relaxation time $\tau$ =  0.50 $\pm$ 0.1 s;
\item Pedestrian radius $R$ = 0.2 $\pm$ 0.02 m.
\end{itemize}
The problem was run repeatedly with an increasing number of pedestrians entering the domain per time unit. The resulting pedestrian density was measured in the passage, as well as the average velocity. The results obtained are compared with published data in figure \ref{fig:velocitydensity-11}. As one can see, the proposed model provides surprisingly accurate results.

\section{Microscopic simulation of the Mataf}
The pilgrimage (called Hajj in Arabic) is the most significant factor in the life and growth of the holy city Mecca. The number of pilgrims to Mecca increased in 2009 by 5 percent compared to 2008, official statistics said. In the last decade the number of pilgrims has been increasing dramatically, every year close to 4 million pilgrims arrive in the Saudi territory through Djedah airport, Djedah port by the Red Sea and through the high way connecting Saudi Arabia with neighbouring states. The arrival of the huge number of pilgrims each year during the last month of the Islamic calendar makes this event a grand human gathering in the world as well as one of the largest logistical and administrative undertakings \cite{Makkah2013}. The goal of pilgrims is to perform the Hajj ritual. Managing the flow of pilgrims to Mecca every year, posed various challenges to Saudi authorities \cite{HajjCommittee2013}.

Millions of believers perform the Hajj at the same time period, 4 specific days in the year. All pilgrims perform the same rituals, typically in same or concurrent time periods. The rituals are in specific points called points of interest, such as the Kaaba in Mecca, the mount Arafat etc. From the mountain Arafat, the pilgrims continue their course to Muzdalifah to throw stones at the pillars symbolizing the devil. Millions of believers live during the Hajj in tents in the Mina area (tent city). After that, millions of believers move towards Mecca to do their final Tawaf Al-Ifadah and to finish the Hajj ritual.

In this section we are concerned with Tawaf as an important pilgrims activity in Hajj period. The accurate prediction of pedestrian motion can be used to assess the potential safety hazards and operational performance at events where many individuals are gathered.
The prediction of the pedestrian flow on the Mataf as permanent by overcrowded area and the investigation of the building facility through the simulation can be used to detect the critical points with high density in different regions of the Mataf. We attempt to determine the average capacity of the Mataf and we study the capacity of the mobile Mataf as possible solution to reduce the pressure of pedestrian flow during the rush hour. As consequence of this solution we investigate how the columns of the mobile Mataf influence the fluidity of pilgrims stream on the ground Mataf. The columns serve as a support of the mobile Mataf. An understanding of how to alter crowd dynamics in Hajj activities would have a significant impact in a number of other scenarios, e.g. during riots or evacuations.  Evacuation from dangerous zones, restrained places or overcrowded buildings, also represent cases where the prediction of pedestrian motion can be used advantageously.
  
\subsection{Motivation}
Now that the world population, particularly the Muslim population, is increasing, the number of pilgrims will also increase. With the huge development in the transportation technology in recent years the number of pilgrims arriving to Mecca grow systematically. It is a fact that the areas of Holy sites, such as Mina, Arafat, Muzdalifah and Mecca are fixed. This has motivated the Saudi government to find an effective and long-term solution. One of these solutions is to evaluate and build the mobile floor for Mataf expansion. The mobile Mataf has a lot of advantages, it can be built in a short time on the Mataf area during the Hajj period, after that it can be rapidly removed.

\subsection{Observations in the simplified model}
One of most important rituals in Hajj is the Tawaf, it consists of circling the Kaaba in the Holy Mosque seven times counter clockwise. Afterwards the Sa`y is performed, the walking between Safa and Marwa in an enclosed, air-conditioned structure.
During Hajj seasons everything in the Mataf area is dense, and we have a compact state. The pilgrims have body contact in all directions and they have no influence on their movement, they are floated by the stream. This forms structures and turbulences in the flow. This turbulence is very well observable in our video recording \cite{dridi2014}. Density and velocity can also be seen.
The following effects could also be observed during the Tawaf:
\begin{itemize}
\item Edge Effects: when the edges of a crowd move faster than the center of the crowd.
\item Density Effects: crowd compression in local areas can imbalance the crowd flow.
\item Shock Waves: propagation of some effects spreading throughout a densely packed crowd.
\item Speed Effects: higher density causes lower walking speeds, with increasing contact area between pilgrims.
\item Group Effects: groups of pilgrims move together and try to keep together all of the time.
\item Break Out Effects: when pilgrims finish the Tawaf, they try to move directly to the edge of the Mataf. The movement of these pilgrims is normally spiral with increasing distance from the center of the Kaaba.
\item Structure Effects: With higher density, turbulences and structures are created within the flow.
\item The average time for Tawaf is ca. 1 hour.
\item While praying in the Haram, there is no Tawaf at all. i.e. density decreases when praying is going to start and rises when praying has finished.
\end{itemize}

\subsection{The Haram mosque building description}

\begin{figure*}
\begin{center}
\includegraphics[width=\linewidth]{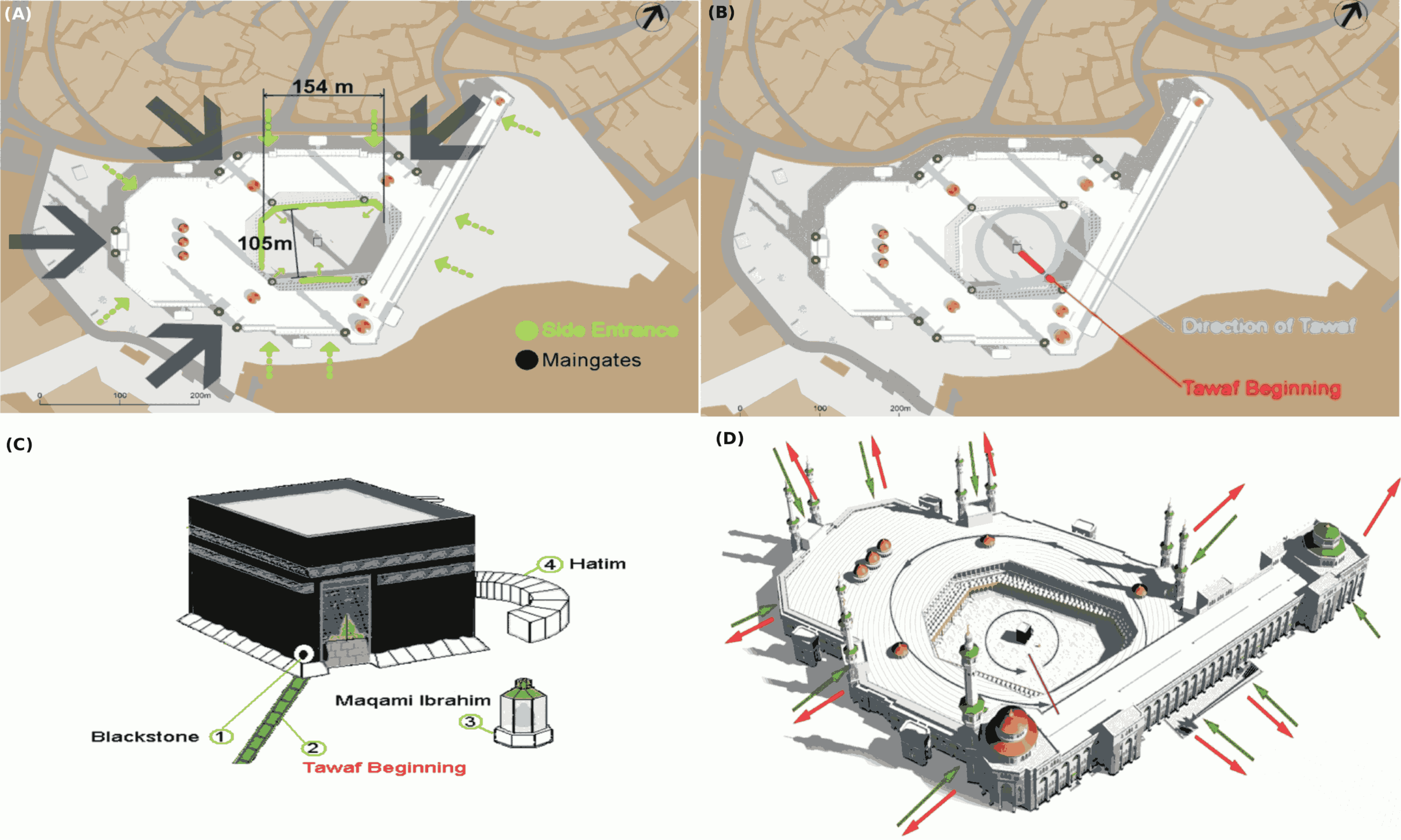}
\end{center}
\caption{(A) Illustration of the Mataf dimension; (B) the direction of the Tawaf movement; (C) the Mataf description: (1) the black stone, one of the most overcrowded regions in the Haram, the measured local density $\rho(\vec{r}, t)$ in this place reached 9 persons/m$^2$, (2) the green strips on the ground indicate the begin and end of the Tawaf movement, (3) Maqam Ibrahim, hysterical monument near the Kaaba that must be considered in the simulation; (D) the gates and entrances to the Mataf.}
\label{fig:haram}
\end{figure*}

The pilgrims stream into the Mataf from all gates of the Haram (see fig. \ref{fig:haram} (A)), but the beginning of Tawaf must start at a specific line (see red line in figure \ref{fig:haram} (B)).  
As we can see in figure \ref{fig:haram} (D), the Haram has 8 gates, stairs and an escalator.
The  gates are probably used by all pilgrims.
Nearly (approximately 80 percent) of pilgrims make Sa'y after Tawaf, the rest (20 percent) of the pilgrims go out through any of the doors.
For the simulation all doors are treated equally.

In Tawaf the pilgrims must circle the Kaaba seven times in a counter-clockwise direction.
The black stone designates the beginning and end of the Tawaf (the pilgrims must stay 5-7 s in front of black stone to appreciate saying "Bismuallah Allahu Akbar" after every round).
Before performing Tawaf, pilgrims try to reach the black stone. They queue around the Kaaba wall with some pilgrims not wanting to queue and trying to push through. 
After the Tawaf people try to pray between Maqami Ibrahim and the Kaaba (see fig. \ref{fig:matafdescription-1}) (The prayer is about 4 to 5 min).  

In figure \ref{fig:haram} (D) a 3D model of the Haram Mosque in Mecca is illustrated. This model shows the main gates, doors, side entrances, and stairs to the Mataf open air area of the Haram. The start/end of the Tawaf is indicated by the red line. The blue line indicates the Tawaf movement direction on the roof of the mosque or in the piazza of the Haram near the Kaaba.

\subsection{Simulation of the simplified model}
\label{sec:simulation_first_set}
\begin{figure*}
\begin{center}
\includegraphics[width=\linewidth]{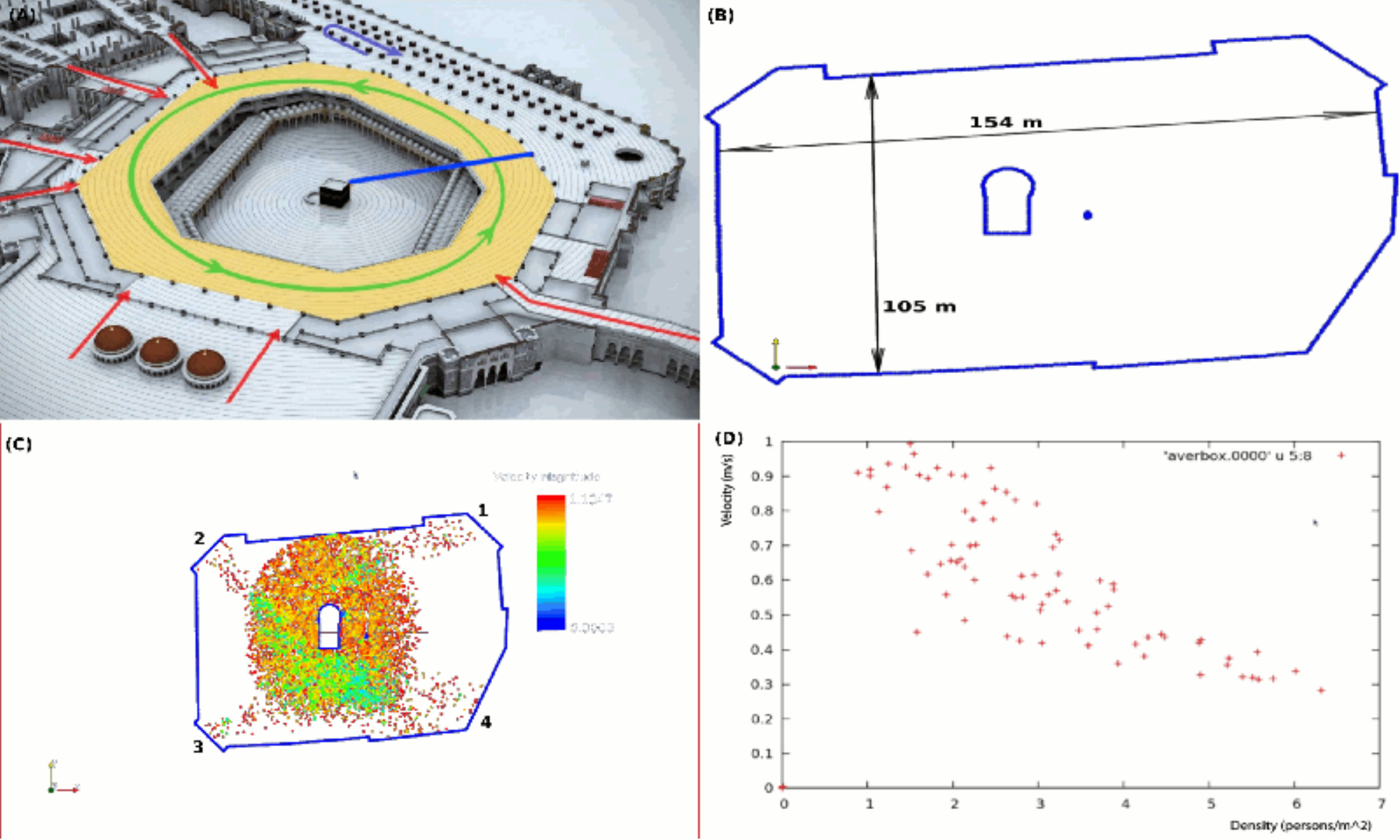}
\end{center}
\caption{(A) 3D-Model of the mosque building in Mecca. Main gates and entrances are indicated through the red arrows, the green arrow indicates the direction of the Tawaf movement, the blue line indicates the beginning and the end of the Tawaf and the pink arrow indicates the Sa'y movement; (B) Mataf dimension; (C) PedFlow simulation snap-shot results, the numbers 1, 2, 3 indicate the entrances or particles streaming in the Mataf area and the number 4 particles streaming out of the Mataf or the exit from the Mataf toward the Sa'y area; (D) decrease of the pedestrian velocity in the Mataf area as a function of the local density, the measured data are represented by the crosses, they are consistent with the empirical data of Predtetschenski and Milinski  \cite{Predtetschenski-Milinski1969}.}
\label{fig:Haramwithoutcoul2}
\end{figure*}

Our first set of simulations consisted of the pedestrian flow through the Mataf area without columns (see fig. \ref{fig:Haramwithoutcoul2} (B)). The Mataf area was 105 m wide and 154 m long and in the center of the Mataf the Kaaba with 21.14 m long and 11.53 m wide is placed. The circumambulation area (ground floor) of Mataf is around 16 170 square meters. Each pedestrian's desired speed was set to a relatively high value of 1.2 m/s, the pedestrian radius $R$ was 0.178 m to 0.2 m and the relaxation time $\tau$ was set in the range of 0.5 s to 1 s. After a pedestrian finishes the Tawaf (seven times circling around the Kaaba) they go in the direction of Sa'y. The measured densities are 5 - 6 persons/m$^{2}$ (measured in different part of the Mataf area with respect to the distance from the Kaaba). These results are consistent with those
obtained by Predtetschenski and Milinski \cite{Predtetschenski-Milinski1969}, who also found out that high density results in reduced velocities. The congested area increases the density sometimes up to 7 or 8 persons/m$^{2}$. In consequence the pedestrians begin to push to increase their personal space and create shock-waves propagating through the crowd.

Figure \ref{fig:Haramwithoutcoul2} (C) displays a pilgrims movement simulation within Mataf area. The entire influx consists of three uni-directional pedestrian flows coming from three entrances. The velocity indicator shows that the movement in the edges of the Mataf area is faster than in the area of the Kaaba. The picture shows a snapshot of the simulation at 38 s, which has a particularly high density of 6.5 persons/m$^{2}$. Also note that the pedestrian density is very high at the places where the Tawaf begins and ends. Note the clumping of pedestrians going in opposing directions, when the pilgrims finish the Tawaf. The average density for many runs was 5 to 6.5 persons/m$^{2}$, (see fig. \ref{fig:Haramwithoutcoul2} (D)).

\subsection{Simulation of the enhanced model}
\begin{figure*}
\begin{center}
\includegraphics[width=\linewidth]{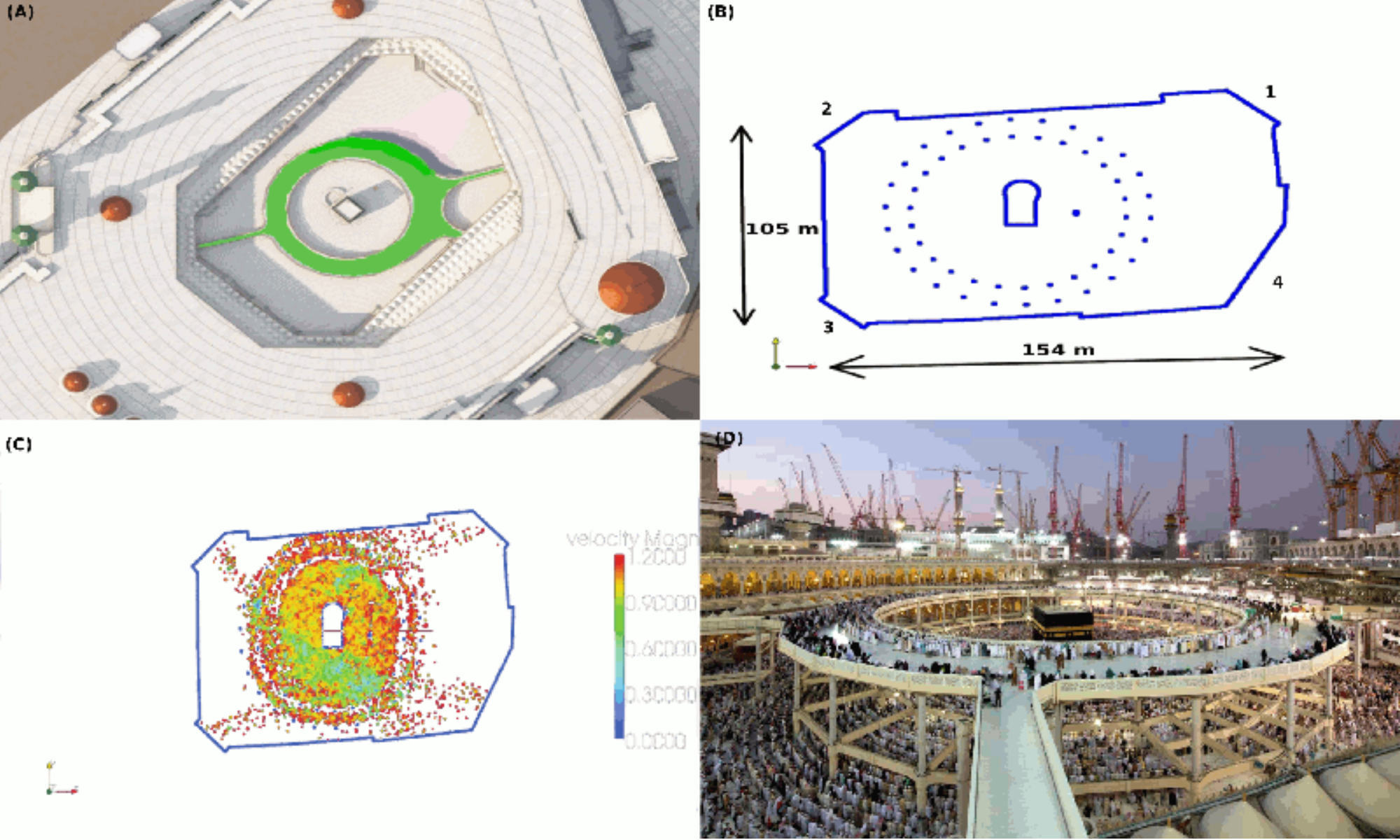}
\end{center}
\caption{Snapshot 1: (A) 3D Model visualises one favourite concept for developing of the Mobile mataf (see green coloured area); (B) CAD drawing displaying the Mataf dimension area with columns; (C) simulation result demonstrates how the pillars influence the movement of pilgrims on the Mataf ground. The colour coding represents the magnitude of the local velocity, PedFlow snapshot; (D) mobile floor for the expansion of the Mataf area in Haram mosque.}
\label{fig:Haramwithcolumn2}
\end{figure*}

The second set of simulations consisted of the pedestrian flow through the Mataf area with columns, (see fig. \ref{fig:Haramwithcolumn2}). The
parameters are the same as in section \ref{sec:simulation_first_set}. After pedestrians finish the Tawaf (seven times circling around the Kaaba) they are going in the direction of Sa'y. The measured densities are again 5 - 6 persons/m$^{2}$ with up to 7 or 8 persons/m$^{2}$ in the congested area. Thus, all results are identical with the simulation without the columns, cf.\ section \ref{sec:simulation_first_set}. Obviously the columns do not influence the densities and velocities of
the pilgrims in the congested area. The only observable difference are the vacant rings on the circles defined by the positions of the columns.

\begin{figure*}
\begin{center}
\includegraphics[width=\linewidth]{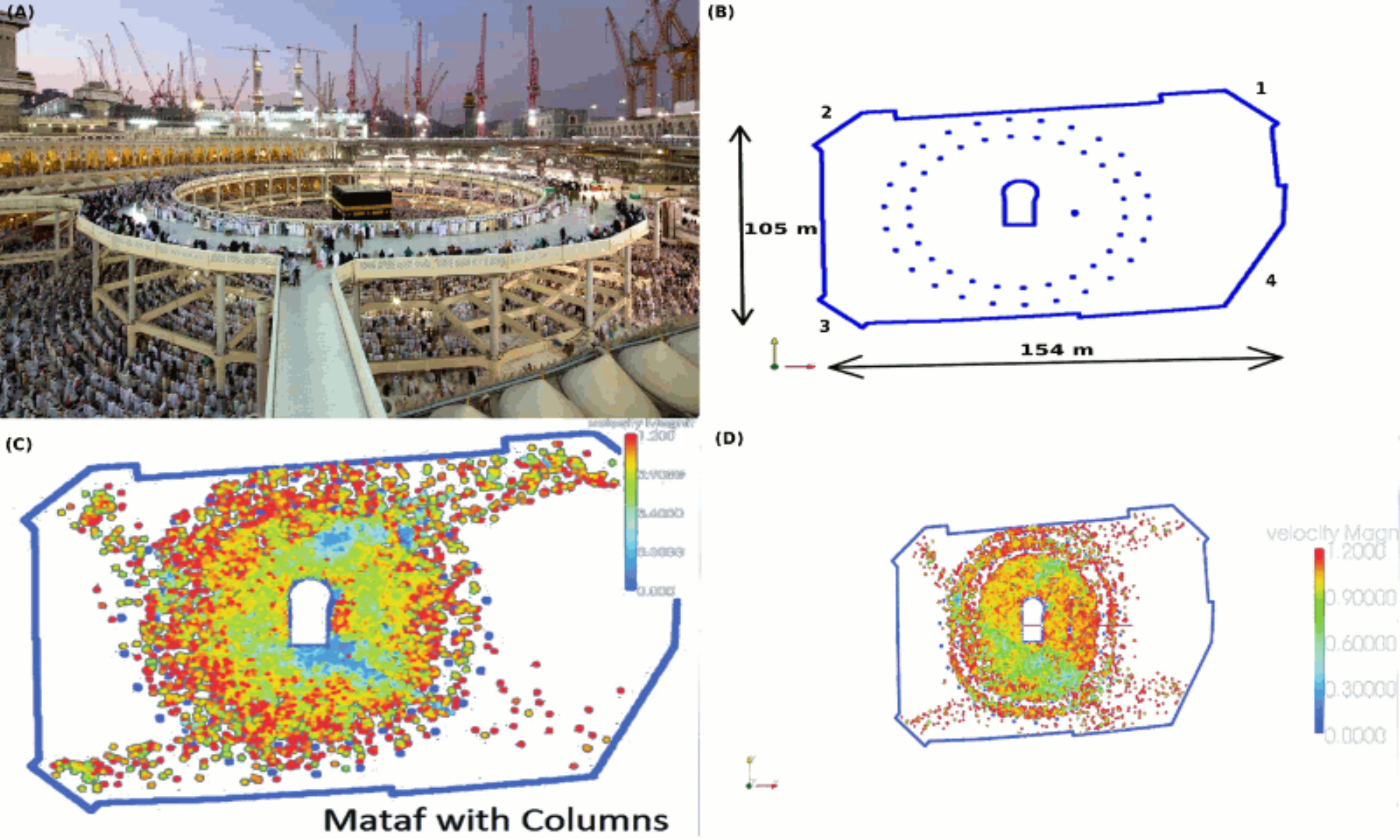}
\end{center}
\caption{Snapshot 2: (A) (Mobile Mataf Floor) at Mataf area in Al-Haram Sharif, one can see the columns (pillars) bearing the mean structure of the temporary Mataf; (B) A CAD drawing displaying the Mataf area with columns; (C) and (D) illustrating a movement of pilgrims in the Mataf area (PedFlow simulation results). The simulation shows a vacant ring since the pilgrims avoid the circle on which the columns are placed. In the region of the beginning and end of the Tawaf ingoing people push back the outgoing people. Thus, it creates a trap becoming more and more crowded.}
\label{fig:Haramwithcolumn3}
\end{figure*}
Figure \ref{fig:Haramwithcolumn3} illustrates two simulation snap-shots with different entire influx. The pedestrian streaming the Mataf area  from three entrances are indicated by the numbers 1, 2, 3 in figure \ref{fig:Haramwithcolumn3} (B). The velocity indicator shows furthermore that the congested area still is in the interior zone near the Kaaba and at the area where the Tawaf begins and ends. The picture shows a snapshot of the simulation at 38 s, which has a particularly high density of 6.5 persons/m$^{2}$. 

For the first moment one can not realise the difference to the results illustrated in figure \ref{fig:Haramwithoutcoul2} and figure \ref{fig:Haramwithcolumn2}, both pictures exhibit the same critical points. Note the clumping of pedestrians going in opposing directions, when the pilgrims finish the Tawaf. One may conclude that the columns have little influence on the movement of pilgrims but one can clearly see a vacant ring along the columns and this results from repulsive interaction between the particles and the columns, this effect is better seen in case of lower crowd density. In case of high density crowd the pressure between the individuals is so huge that this vacant ring along the column disappears. This means that under the enormous surrounding pressure, people are forced to come near the pillar. Here we stress that the small size and smooth form of the column is very important to prevent people injury or dangerous situations. 

\begin{figure*}
\begin{center}
\includegraphics[width=\linewidth]{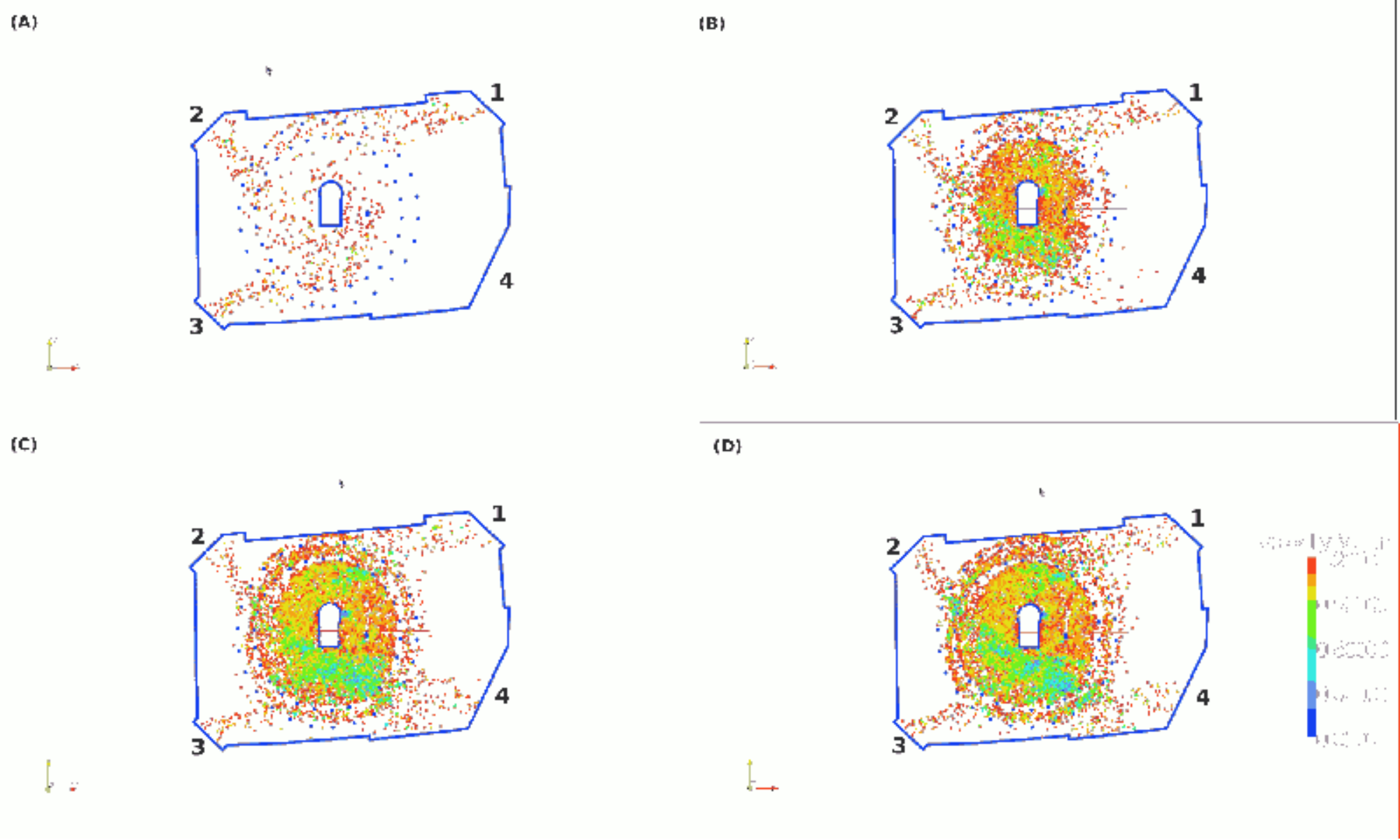}
\end{center}
\caption{Pilgrims streaming the Mataf area through three different gates denoted by 1, 2, 3 and leaving the Mataf floor through gate 4. Step by step simulation of a pedestrian flow circling seven time the holy Kaaba  ((A), (B), (C) and (D) illustrate time consecutive PedFlow animation snapshots). The impact of the columns on the pilgrims movement performing the Tawaf are demonstrate. The red colour indicates pedestrian walking with their desired velocities and the turquoise colour indicates pedestrian walking with lower velocities.}
\label{fig:mmataf7}
\end{figure*}

\section{Mataf capacity estimation}
\subsection{Fundamental Diagram}
\begin{figure}[H] 
\begin{center}
\includegraphics[width=\columnwidth]{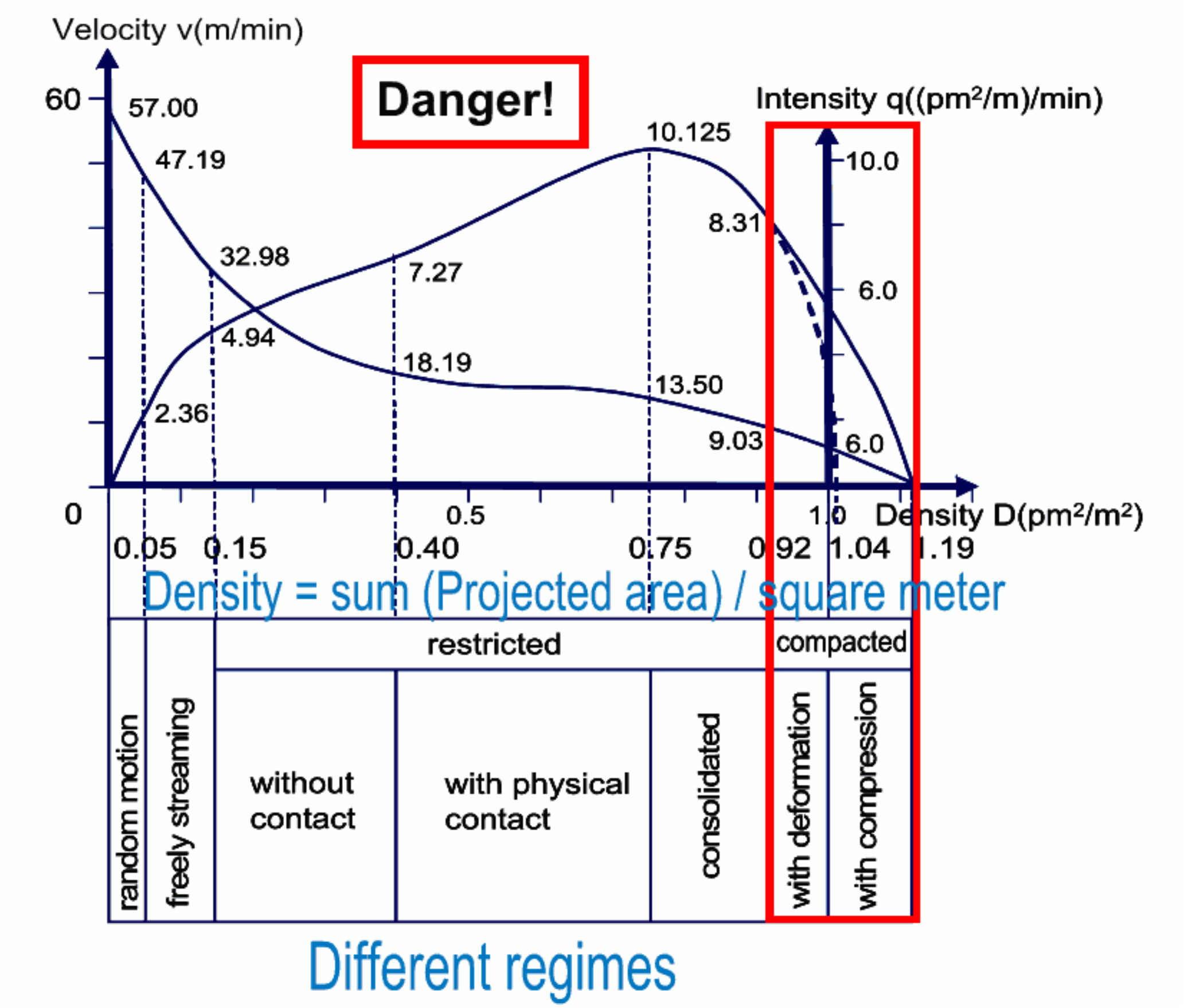}
\end{center}
\caption{Velocity-density diagram; Empirical relation between density and velocity according to Predtetschenski and Milinski \cite{Predtetschenski-Milinski1969}. The partition refers to domains with qualitatively different decrease of the velocity. The red rectangle indicates a dangerous area in this curve, where the pedestrian density reaches high values with more than 6 persons/m$^{2}$. In this range bodies will be pressed or deformed, resulting in high injury scales with possible fatal consequences.}
\label{fig:fundamentaldiagram-1}
\end{figure}

Pedestrian distribution on the used surface must be identified with the fundamental diagrams (see fig. \ref{fig:fundamentaldiagram-1}).

\subsection{Calculations}
We divided the Mataf area into three parts according to the density distribution.
The first part is the area near the Kaaba wall.
The second part is the middle of the Mataf, the area between 10 m and 20 m from the Kaaba wall.
The third part is the edge of the Mataf, the area between 20 m and 30 m.

\begin{itemize}
\item The first part of the Mataf area:
\begin{itemize}
\item Usable Area = 820 m$^{2}$.
\item The mean pedestrian density = 6 pilgrims/m$^{2}$. 
\item The total number of pilgrims = 4920 pilgrims.
\item The mean pedestrian path = 91 m.
\item The mean Tawaf path = 91 $\times$ 7 = 637 m.
\item The Tawaf lasts 39 min with average walking velocity of 0.266 m/s = 15,96 m/min.
\item The capacity is 7596 pilgrims Tawaf/hour    
\end{itemize}
\end{itemize}
\begin{itemize}
\item The second part of the Mataf area 
\begin{itemize}
\item Usable Area = 1380 m$^{2}$.
\item The mean pedestrian density = 3,68 pilgrims/m$^{2}$.
\item The total number of pilgrims = 5078 pilgrims.
\item The mean pedestrian path = 151 m. 
\item The mean Tawaf path = 151 $\times$ 7 = 1057 m.
\item The Tawaf takes 52,85 min with average walking velocity of 0,33 m/s = 20 m/min.
\item The capacity is 5859 pilgrims Tawaf/hour.
\end{itemize}
\end{itemize}
\begin{itemize}
\item The third part of the Mataf area
\begin{itemize}
\item Usable Area = 1900 m$^{2}$.
\item The mean pedestrian density = 2 pilgrims/m$^{2}$ .
\item The total number of pedestrians = 3800 pilgrims.
\item The mean pedestrian path = 212 m. 
\item The mean Tawaf path = 212 $\times$ 7 = 1484 m.
\item The Tawaf lasts 49,46 min with average walking velocity of 0.5 m/s = 30 m/min.
\item The capacity is 4653 pilgrims Tawaf/hour.
\end{itemize}
\end{itemize}
The total capacity of the Mataf is 18923 pilgrims Tawaf/hour.

According to several studies about the Haram, the actual capacity of the entire Mataf (ground, intermediate floors and the roof) are around 50,000 pilgrims/hour, but the capacity of the new Massaa is much higher than the capacity of the existing Mataf. During peak times worshippers are in danger of injuries due to extreme high density in the Mataf. To increase the capacity and improve the conditions for worshippers a new additional floor for the Mataf called temporary Mataf has been developed.

\section{Mobile Mataf geometry}
The new floor is based on a modules concept with fast assembly and disassembly without interruption of ongoing Tawaf. This light weight structure is designed to be integrated into the existing Haram Mosque. The outer diameter is about 94 m, the inside diameter is around 70 m and the width is around 12 m (see fig. \ref{fig:mmataf5} (A)). An additional area of 3000 m$^{2}$ is
added to the existing Mataf area.
Access is achieved over several ramps. The clearance height is around 2.7 m leaving enough space for worshipper to pass underneath.

\subsection{Concepts and evaluations}
The idea to develop a new moving platform above the existing Mataf has the goal to take pedestrian pressure from the old Mataf. The design and evolution of this mobile area revealed a new concept concerning the safety and fluidity of pilgrims flow. In the next paragraphs we discuss two approaches with two different geometries and we present our analysis data.

\begin{itemize}
\item Mobile Mataf with extension: we shall discuss
\begin{itemize}
\item Geometry
\item Simulation results 
\item Capacity of the Mataf according to Predtetschenski and Milinski densities
 \end{itemize}
 \end{itemize}

\subsection{Simulation}
This set of simulations considers a unidirectional pedestrian flow circling the Holy Kaaba seven times. Different strengths of the influx of pilgrims were used, see table \ref{tab:simulation-1}.The walking pedestrian area named mobile floor for the expansion of the Mataf contains two exits and one emergency exit. The design of the Mobile Mataf allows one-directional pedestrian flow (see fig. \ref{fig:mmataf5} (A)). The parameters used in the simulations were the desired velocity of the individual $v_{d}$, the individual radius $R$ and the relaxation time $\tau$. The size or radius of particles are given in Table \ref{tab:projectionarea} and according to Predtetschenski and Milinski \cite{Predtetschenski-Milinski1969}. For simplicity, all  
parameters are identical for each person.

\begin{table}[H]
\begin{center}
 \small
\begin{tabular}{|p{4cm}|p{2cm}|  }
\hline
Type of Person & Projected area (m$^{2}$) \\ 
 \hline
 Children & 0.04 \textendash  0.06 \\ 
 Adolescent & 0.06 \textendash  0.09 \\ 
 Grown-up in summer clothes & 0.100 \\ 
Grown-up in interseason & 0.113 \\ 
 Grown-up in winter clothes  & 0.125 \\ 
 Grown-up in interseason clothes with briefcase & 0.180 \\ 
 Grown-up in interseason clothes with light luggage & 0.240 \\ 
 Grown-up in interseason clothes with heavy luggage & 0.390 \\
  \hline  
\end{tabular}
\end{center}
\normalsize
\caption{Set of different individual sizes \cite{Predtetschenski-Milinski1969}.}
\label{tab:projectionarea}
\end{table}

\subsubsection{Parameters}
\begin{itemize}
\item Individual Parameters.
\begin{itemize}
\item Desired velocity: $v_{d}$ = 1.2 $\pm$ 0.2 m/s.
\item Individual Radius: $R$ = 0.178 $\pm$ 0.02 m (Milinski).
\item Relaxation Time: $\tau$ = 0.5 $\pm$ 0.1 s.
\end{itemize}
\item Geometry Parameters.
\begin{itemize}
\item Usable Area: $A$ = 3540 m$^{2}$.
\item The mean Radius of the Mataf: $R_{M}$ = 42 m.
\item The mean pedestrian path: $L_{Path}$ = $2 \pi R_{M}$ = 263.76 m.
\item The mean Tawaf path: $L_{Tawaf}$ = 263.7 $\times$ 7 = 1846.32 m.
\item The Tawaf time: $t_{Tawaf}$ = 43.35 min with average walking velocity $\bar{v}$ =  0.7 m/s = 42 m/min.
\end{itemize}
\end{itemize}

\subsubsection{Results}

\begin{table}[H]
\centering
 \tiny
\begin{tabular}{|p{1cm}||p{1cm}||p{1cm}||p{1cm}||p{1cm}|}
\hline { Influx (Persons/s)} & { Max Number of Ped} & { Density(1/m$^{2})$} & { Capacity(Tawaf/h)} & { Remarques} \\ 
\hline 2,0 & 5032 & 1.42 & 7021 & fluid \\ 
\hline 2,1 & 5338 & 1.50 & 7448 & fluid \\ 
\hline 2,2 & 5796 & 1.63 & 8087 & fluid \\ 
\hline 2,3 & 6147 & 1.73 & 8577 & fluid \\ 
\hline 2,4 & 6691& 1.89 & 9336 & fluid \\ 
\hline 2,5 & 6955 & 1.96 & 9704 & fluid \\ 
\hline 2,6 & 7443 & 2.10 & 10385 & fluid \\ 
\hline 2,7 & 8106 & 2.28 & 11310 & fluid \\ 
\hline 2,8 & 8519 & 2.40 & 11886 & fluid \\ 
\hline 3,9 & 9236 & 2.60 & 12887 & Slow Movement \\ 
\hline \color{red}3,0 & \color{red}10358 & \color{red}2.92 & \color{red}14453 & \color{red}{ Clogging effect}\\ 
\hline
\end{tabular}
\normalsize
\caption{The Mataf capacity at different simulation influxes.}
\label{tab:simulation-1}
\end{table}

In table \ref{tab:simulation-1} the maximum number of pedestrians on the Mataf without clogging is 8404 with a density of 2.37 persons/m$^{2}$. 
The optimal capacity without clogging is 11726 pilgrims per hour completing the Tawaf, calculated with an average Tawaf duration of 43 min. 

\begin{figure*}
\begin{center}
\includegraphics[width=\linewidth]{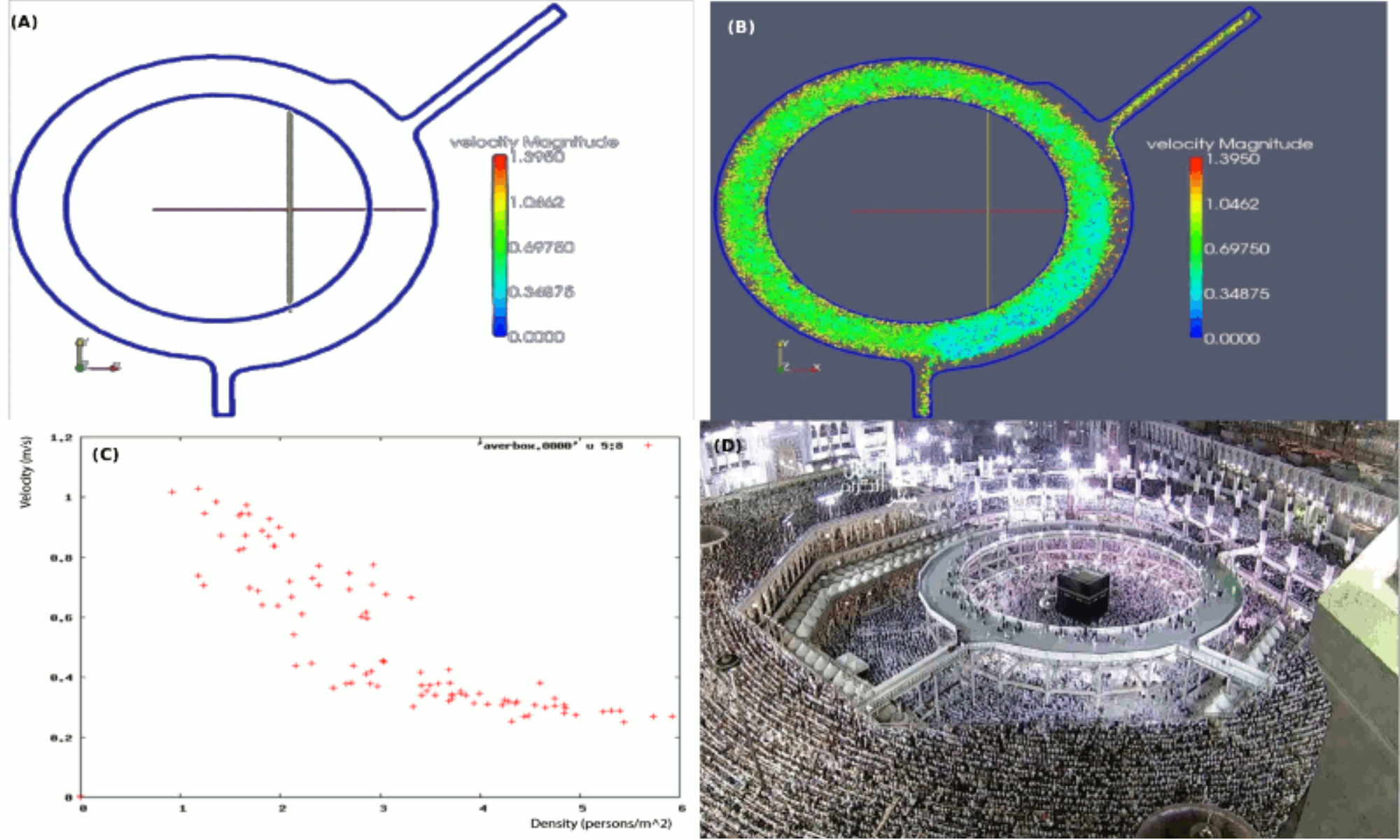}
\end{center}
\caption{(A) CAD drawing displaying one of many geometry concepts developed for evaluation of the mobile Mataf; (B) simulation results (new geometry concept); (C) velocity density diagram (PedFlow simulation); (D) the actual mobile floor facility top view.}
\label{fig:mmhauggeometrie-1}.
\end{figure*}

The maximum number of pedestrians on the Mataf was 11629 with a density of 2.79 persons/m$^{2}$. The optimal capacity without clogging was 16226 Tawaf/hour, calculated with an average Tawaf duration of 43 minutes.
If one Tawaf is 50 minutes then the capacity of the Mataf will be 13954 Tawaf/hour.

\section{Mataf capacities estimation}
\begin{figure*}
\begin{center}
\includegraphics[width=\linewidth]{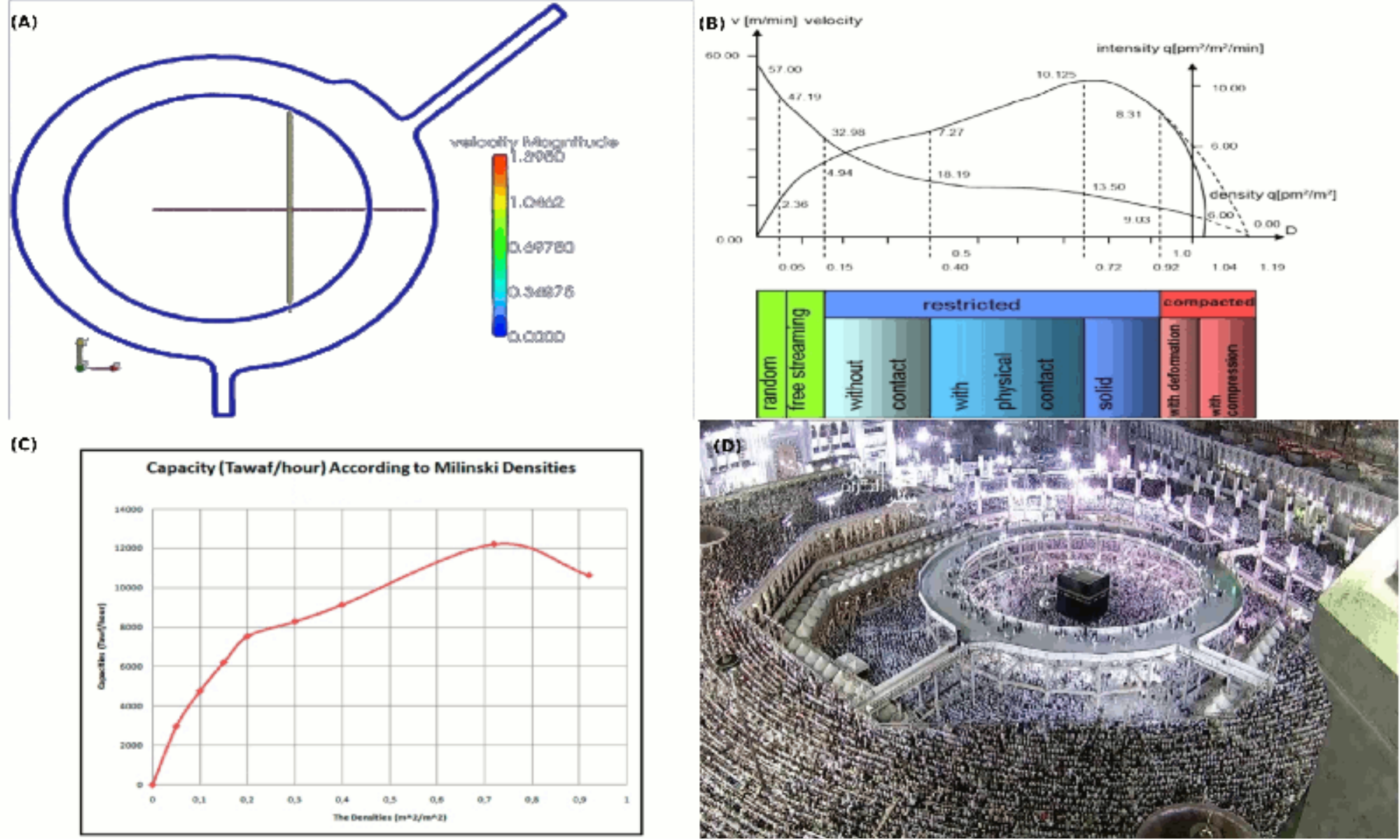}
\end{center}
\caption{(A) CAD drawing displaying the geometry concept of the new walkway around the Kaaba; (B) Predtetschenski and Milinski fundamental diagram; (C) mobile floor capacity according to Predtetschenski and Milinski densities; (D) the present mobile floor facility top view.}
\label{fig:mmataf5}
\end{figure*}

According to the Predtetschenski and Milinski fundamental diagram figure \ref{fig:fundamentaldiagram-1}, the capacity of the mobile Mataf was determined. The results of this investigation are illustrated in Table \ref{tab:matafcapacity-3.}. The capacities of the mobile floor were computed with respect to different pedestrian densities. Every density in the fundamental diagram is related to a certain velocity. With help of these given velocities we tried to calculate the mean average time for one complete Tawaf. Afterwards we calculated the capacity of the Mataf. We can see that the optimal capacity is about 12000 Tawaf/hour, (see fig. \ref{fig:mmataf5} (C)). This result consistent with the simulation result which is about 13954 pilgrims per hour completing the Tawaf. 

\begin{table}[H]
\centering
 \tiny
\begin{tabular}{|p{0.5cm}||p{0.5cm}||p{0.5cm}||p{0.5cm}||p{0.5cm}||p{0.5cm}||p{0.5cm}|}
\hline Density (1/m$^{2})$ & Velocities (m/min) & Path length (m) & Tawaf time (min) & Usable Area
(m$^{2})$ & Max Pedestrian number & Capacity (Tawaf/hour) \\ 
\hline 0 & 60 & 1846 &  & 3864 & 0 & 0 \\ 
\hline 0.05 & 47.19 & 1846 & 39.11 & 3864 & 1932 & 2963 \\ 
\hline 0.1 & 38 & 1846 & 48.57 & 3864 & 3864 & 4773 \\ 
\hline 0.15 & 32.98 & 1846 & 55.97 & 3864 & 5796 & 6213 \\ 
\hline 0.2 & 30 & 1846 & 61.53 & 3864 & 7728 & 7535 \\ 
\hline 0.3 & 22 & 1846 & 83.90 & 3864 & 11592 & 8289 \\ 
\hline 0.4 & 18.19 & 1846 & 101.48 & 3864 & 15456 & 9138 \\ 
\hline \color{red}0.72 & \color{red}13.5 & \color{red}1846 & \color{red}136.74 & \color{red}3864 & \color{red}27820 & \color{red}12207 \\ 
\hline 0.92 & 9.03 & 1846 & 200.65 & 3864 & 35548 & 10629 \\ 
\hline 
\end{tabular} 
\normalsize
\caption{Mataf capacities according to Predtetschenski and Milinski densities.}
\label{tab:matafcapacity-3.}
\end{table}

\section{Recommendations for the future Hajj planning and development}
With the increasing number of pilgrims every year and with the dramatic evolution of the transportation system in the world, the design and architecture of the Haram mosque in Mecca must be changed to
increase the capacity, throughput and performance.
The proposed and evaluated design has a specific aim, the safety of pilgrims. In order to ensure that the proposed design satisfies the required criteria of pilgrims safety, especially during overcrowding, many simulations and tests were realized to access the critical points in the Mataf area. The new Mataf area was designed to satisfy the international standard criterion of pedestrians safety, especially during overcrowding. The flow of pilgrims is demanding, since any accumulation in this area can be dangerous and has fatal consequences.
The new proposed design offers a significant improvement to the safety of pilgrims. The limited spatial possibilities in and around the Haram area for infrastructural development requires an in-depth analysis and search for solutions to solve the problem of pedestrian flow. From this the concept arises to build a temporary Mataf to take a small capacity from the ground Mataf as best arrangements during the massive extension project in Masjid Al-Haram.

\begin{figure}[H]
\begin{center}
\includegraphics[width=\columnwidth]{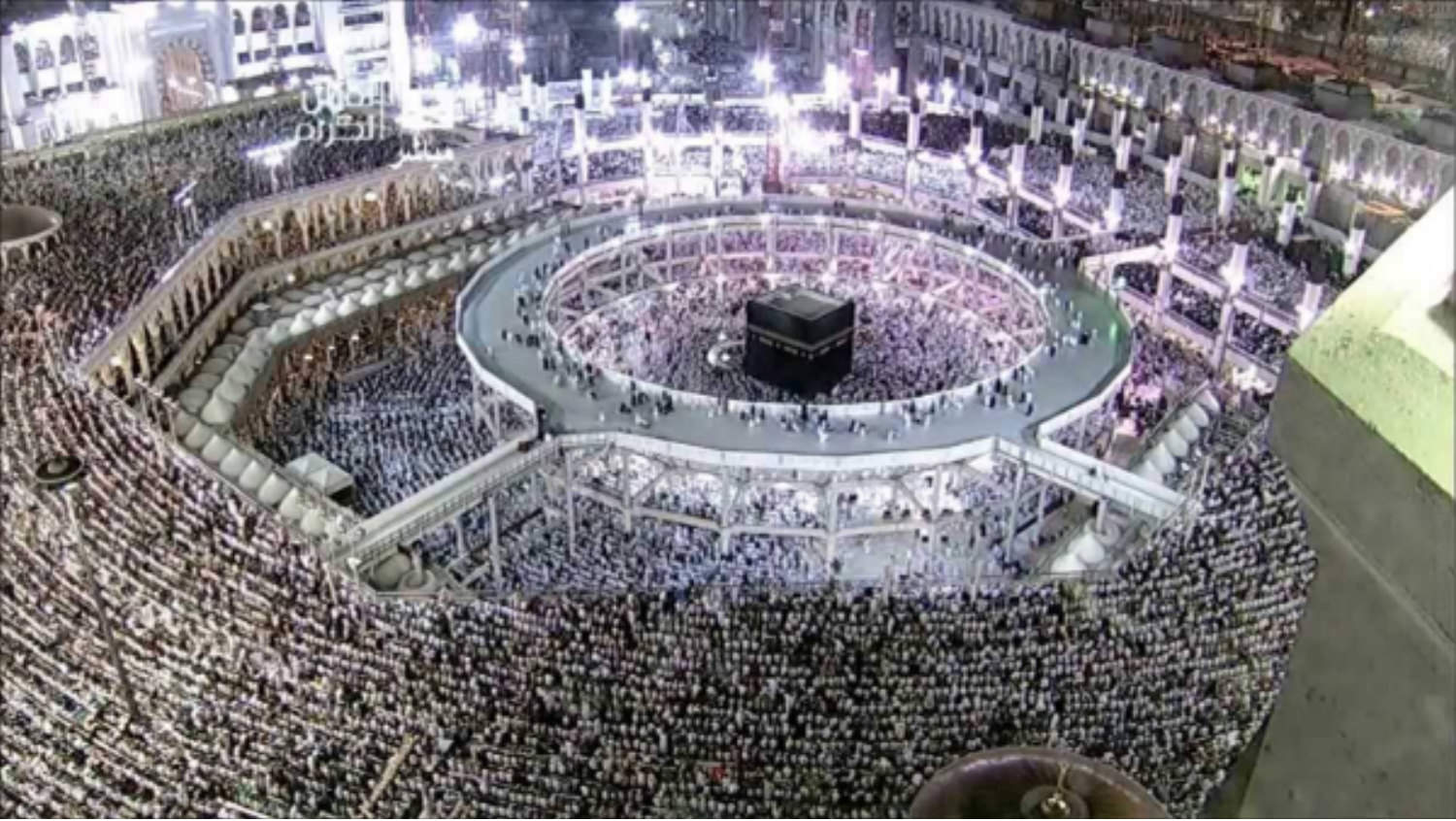}
\end{center}
\caption{The new walk way around the Kaaba is definitely taking shape 
in this picture (26.07.2013).}
\label{fig:mobile-mataf-extension}
\end{figure}

The Mataf area is a very restrained plain. The construction of the temporary Mataf within Masjid Al Haram should attempt to resolve many problems to avoid a potential disaster in the future, such as the overcrowding disaster at the Jamarat bridge in 2005.
The conceptual design as we can see in figure \ref{fig:mobile-mataf-extension} has two problems; the first problem concerns the fluidity of the movement on the platform especially at the exits, the second problem concerns the ground Mataf in that the fluidity of pedestrian flow will be disturbed through the pillars carrying the platform.
The Haram building must ensure the international norm of safety to allow the increased number of pilgrims.

To prevent stampede and congestion within the Mataf area during the rush hour of the Hajj events a sufficient study and analysis of the expected inflows and outflows (and, hence, number of pilgrims) must be ascertained and documented, considering the alternation possibility of large pedestrian flows. A narrow passage and bottleneck analysis within the Haram mosque area is crucial for any decision making. Through exact simulation and analysis of pedestrian fluidity on the Haram area the congestion points as the intersection or crossing points must be established and should be removed. Either the passages with counter flow must be detected as early as possible and carefully treated (examined). This analysis can be used to assess the potential safety hazards and can be helpful for the organizers.
As critical points the exits of the Haram must be checked. The passage between the Mataf area and the Massaa area and between the pillars of the temporary Mataf can cause problems for the fluidity of the pilgrim streams.

For the safety of all pilgrims on the Haram area and to ensure a quick and
effective response to any occurring problems a sophisticated and continuously monitoring system (e.g. video surveillance) must be established.

An engineering system must be developed within the Haram to allow the security and emergency forces to move as quickly as possible and unobstructed to any place if something happens, to remove or at least mitigate problems.

The main problem in Mecca itself are the traffic jams. The high density of cars and buses combined with walking or road crossing pedestrians make the situation more complex for rescue or evacuation plans.
The possibility to be stampeded at Hajj pilgrimage is very high at all places and especially inside the Haram in Mecca, in Muzdalifa, at the Jamarat and on the streets between the so called points of interest. 

The changes necessary for a successful Hajj and its activities are all contained in the wider long-term plan for Mecca. The next expansion for Haram and Massaa requires a new infrastructure and regeneration of the east end - which will reclaim old neighbourhoods to Haram for housing and parking area - which is a huge project.

\section{Conclusions and possible enhancements}
A newly developed simulation software for analysing pilgrim movement was used for the evaluation of the Tawaf capacity of the Mataf area and the new mobile floor. Different scenarios were analysed to find out the optimal capacity. With a flux of only 12 pilgrim/min entering the floor, much space was left unused. Pilgrims moved with their desired speed and the average time for completing a Tawaf took around 25 minutes.
The second scenario showed 108 pilgrim/min entering the floor, causing the space to become fully occupied by pilgrims and
the velocity to decrease, resulting in dangerous jams especially in the exit area.
The third scenario showed the optimum steady state when about 125 pilgrim enter per minute and also exit the floor, with no jams. This results in an overall capacity of about 9000 finished Tawaf/hour and an average time of approximately 25 min to complete the Tawaf.

This work is an attempt to model high density crowd dynamic flows in autonomous and controlled scenarios. A microscopic approach was used to model this problem based on observed collective behaviour in emergency conditions, where the detailed design of interactions is overlapped by group behaviour. We presented a simple means of achieving more complex behaviour in both indoor and outdoor environments. The simulation tools produce results that compare favourably with the real data. 
 
As a practical example, the Haram Mosque in Mecca and the Jamarat Bridge in Saudi Arabia were used for high density crowd simulation: the huge number of pilgrims cramming the bridge during the pilgrimage to Mecca caused serious pedestrian disasters in the 90's. Moreover, the analytical and numerical study of the qualitative behaviour of human individuals in a crowd with high densities can also improve traditional socio-biological investigation methods. 

For obtaining empirical data different methods were used, automatic and manual methods. We analysed video recordings of the crowd movement in the Tawaf within the mosque in Mecca during the Hajj of the 27th of November, 2009. We evaluated unique video recordings of a 105$\times$ 154 m large Mataf area taken from the roof of the Mosque \cite{dridi2014}.

For the validation and calibration of the simulation tools, different methods were used \cite{dridiv2014}. 
At medium to high pedestrian densities, the techniques used in PedFlow can produce realistic crowd motion, with pedestrians moving at different speeds and under different circumstances, following believable trails and taking sensible avoidance action.

As possible future work, we would like to have a better understanding of the parameters influencing the crowd behaviour such as the ethnic and psychological parameters. These parameters have a significant impact on the movement of crowds especially in panic situations.
The psychological-physiological parameters in the case of riots or emergencies must be investigated further and documented so that they can be implemented into other models.  
Our proposed approach to simulate crowd behaviours in the future is
based on this dichotomy, and we propose separate models for an agent's personality and another one to account for situational factors.
This question arises and opens the door to many topics that must be further investigated. There are certain things we can do to help a high density crowd participant like pilgrims, sport or music event spectators to survive a stressful situation, by planning, preparation, training and better management. But this is not sufficient to make sure that hundred percent of the crowd are safe, if we consider the number of accidents happened in the last decade.

\section{Acknowledgements}
I would like to express my sincerest thanks and gratitude to Prof. Dr. G. Wunner for a critical reading of the 
manuscript, for his important comments and suggestions to improve the manuscript. Many thanks to Dr. H. Cartarius for his support during writing this work.


\end{multicols}
\end{document}